\documentclass[preprint]{aastex}
\pdfoutput=1

\newcommand{\gsim}{\mbox{$\stackrel {>}{_{\sim}}$}} 
\newcommand{\lsim}{\mbox{$\stackrel {<}{_{\sim}}$}}

\slugcomment{Accepted to ApJ}

\shorttitle {$^{13}$CO and C$^{18}$O in Maffei 2}
\shortauthors{Meier et al.}

\begin{document}

\title{Nuclear Bar Catalyzed Star Formation: $^{13}$CO, C$^{18}$O and
Molecular Gas Properties in the Nucleus of Maffei 2}

\author{David S. Meier\altaffilmark{1,2,3}, Jean L. Turner\altaffilmark{4}, 
and Robert L. Hurt\altaffilmark{5}}

\altaffiltext{1}{David S. Meier is a Jansky Fellow of the National 
Radio Astronomy Observatory}
\altaffiltext{2}{National Radio Astronomy Observatory,
P. O. Box O, 1003 Lopezville Road, Socorro, NM 87801; email: dmeier@nrao.edu}
\altaffiltext{3}{Department of Astronomy, University
of Illinois, 103 W. Green St., Urbana, IL, 61801}
\altaffiltext{4}{Department of Physics and Astronomy, UCLA, Los Angeles,
CA 90095-1547; email: turner@astro.ucla.edu}
\altaffiltext{5}{Infrared Processing and Analysis Center, California 
Institute of Technology, MS 100-22, Pasadena, CA 91225; email: 
hurt@ipac.caltech.edu }

\begin{abstract}
We present $\sim3^{''}$ resolution maps of CO, its isotopologues, and
HCN from in the center of Maffei~2. The J=1--0 rotational lines of
$^{12}$CO, $^{13}$CO, C$^{18}$O and HCN, and the J=2--1 lines of
$^{13}$CO and C$^{18}$O were observed with the OVRO and BIMA arrays.
The lower opacity CO isotopologues give more reliable constraints on
H$_{2}$ column densities and physical conditions than optically thick
$^{12}$CO.  The J=2--1/1--0 line ratios of the isotopologues constrain
the bulk of the molecular gas to originate in low excitation,
subthermal gas.  From LVG modeling, we infer that the central GMCs
have $n_{H_{2}}\sim 10^{2.75}~cm^{-3}$ and T$_{k}~\sim 30$ K.
Continuum emission at 3.4 mm, 2.7 mm and 1.4 mm was mapped to
determine the distribution and amount of \ion{H}{2} regions and dust.
Column densities derived from C$^{18}$O and 1.4 mm dust continuum
fluxes indicate the standard Galactic conversion factor overestimates
the amount of molecular gas in the center of Maffei~2 by factors of
$\sim$2-4.  Gas morphology and the clear ``parallelogram'' in the
Position-Velocity diagram shows that molecular gas orbits within the
potential of a nuclear ($\sim$220 pc) bar.  The nuclear bar is
distinct from the bar that governs the large scale morphology of
Maffei~2.  Giant molecular clouds in the nucleus are nonspherical and
have large linewidths, due to tidal effects.  Dense gas and star
formation are concentrated at the sites of the $x_{1}-x_{2}$ orbit
intersections of the nuclear bar, suggesting that the starburst is
dynamically triggered.
\end{abstract}

\section{Introduction}

Concentrations of molecular gas are common in the centers of large
spiral galaxies, often in the form of nuclear bars.  Bars represent
likely mechanisms by which rapid angular momentum loss and gas inflow
concentrate gas in the centers of galaxies
\citep[eg.][]{SOIS99,SVRTT05,K05}.  Secondary nuclear bars can be
important in controlling the dynamics of the innermost few hundred
parsecs of galaxies \citep[eg.,][]{SFB89,FM93,HSE01,MTSS02,SH02,ES04}.
In addition to being an avenue by which nuclear gas and stellar masses
are built up, these bars can also significantly influence the physical
and chemical properties of molecular clouds within them
\citep[eg.][]{MT01,PW03,MT04,MT05}.

We have observed the two lowest rotational lines of $^{13}$CO and
C$^{18}$O, plus new high resolution images of $^{12}$CO(1-0) and
HCN(1-0) in the nearby galaxy, Maffei 2, with the Owens Valley
Millimeter (OVRO) Array and the Berkeley-Illinois-Maryland Association
Array (BIMA).  Because they are optically thin, or nearly so, CO
isotopologues more directly trace the entire molecular column density
than does optically thick $^{12}$C$^{16}$O (hereafter ``CO'').  Also
ratios amongst isotopologues are more sensitive to changes in density
and temperature throughout the clouds.  HCN, which has a higher
critical density than CO, constrains the dense molecular cloud
component.  Here we aim to characterize the properties of molecular
clouds in the center of this barred galaxy, and to connect those
properties with the dynamics of the nucleus.

Maffei 2 is one of the closest large spirals (D$\simeq$3.3 Mpc, $\S$2;
Table \ref{Tmaf}), but lies hidden behind more than 5 magnitudes of
Galactic visual extinction \citep[][]{M68}.  A disturbed, strongly
barred spiral galaxy \citep[][]{HMGT93a,BM99}, Maffei 2 has an asymmetric
HI disk and tidal arms that suggest a recent interaction with a small
satellite \citep[][]{HTH96}.  The interaction may be responsible for
the bright nuclear CO emission \citep[eg.,][]{RTP77,I89,MW04} and
active nuclear star formation of L$_{OB}~\simeq ~1.7\times
10^{9}~$L$_{\odot}$ \citep[][corrected for distance]{TH94}.

\section{Distance to Maffei 2}

Galactic extinction complicates distance determinations for this
nearby ($V_{LSR}$=-30 km s$^{-1}$) galaxy.  Some have suggested that
Maffei 2 is close enough to have a significant dynamical influence on
the Local Group \citep[][]{BM83,ZVB91}.  Estimated distances to the
members of the IC 342/Maffei 2 group range from 1.7--5.3 Mpc
\citep[][]{BM83,M89,LT93,KT93,KT94,KTG95,KDKTHV97,IARM99,DvB01}.
Closer distances (D$_{mpc}~\sim$ 2) tend to come from the
Faber-Jackson relationship and the brightest supergiants method
\citep[][]{BM83,KT93,KT94,KDKTHV97}, while the farther distances
(D$_{mpc}~\sim$ 4--5) come from Tully-Fisher relations on Maffei 2 and
surface brightness fluctuations methods towards its companion Maffei 1
\citep[][]{H93,LT93,KTG95,IARM99}.  In several cases, the same method
gives wide ranges of distances for different galaxies within the same
group, suggesting the group is spatially extended (likely) or each
distance measurement has higher uncertainties than claimed
\citep[][]{KTG95,KDKTHV97}.  Recent studies appear to be converging to
D $\sim$ 3-3.5 Mpc for the IC 342/Maffei 2 group
\citep[][]{SCH02,FMDKKLB03,KSDG03,Kar05,FLMR07}.  \citet[][]{FLMR07}
do a self consistent analysis of several different measurements and
obtain a distance of 3.3 Mpc.  We adopt this distance for Maffei 2,
with uncertainties of $\sim$ 50 \%.  Quoted uncertainties in this
paper do not include this systematic uncertainty.

\section{Observations}

Aperture synthesis observations of the $^{13}$CO(1-0), $^{13}$CO(2-1),
C$^{18}$O(1-0), C$^{18}$O(2-1) and HCN(1-0) lines were obtained with
the Owens Valley Radio Observatory (OVRO) Millimeter Array between
1993 October 26 and 1999 March 29.  The $^{13}$CO(1-0) and
$^{13}$CO(2-1) data were obtained when the OVRO array had five 10.4
meter antennas, while the remaining data are from the six-element
array \citep[][]{OVRO91,OVRO94}.  Observing parameters are listed in
Table \ref{Tobs}.  Separate 1 GHz bandwidth continuum channels at 3.4
mm, 2.7 mm and 1.4 mm were also recorded.  3C84 and 0224+671 were used
to calibrate instrumental amplitudes and phases.  Absolute fluxes were
calibrated using Neptune, Uranus and 3C273 as standards, with
additional observations of 3C84 and 3C454.3 as consistency checks.
Absolute flux calibration should be good to 10 -- 15\% for the 3 mm
data and 20 -- 25\% for the 1 mm data, and internally consistent
between each transition.

We have also obtained high resolution ($\sim 3^{''}$) observations of
the $^{12}$CO(1-0) transition with the ten element
Berkeley-Illinois-Maryland Association (BIMA) Array\footnote{Operated
by the University of California, Berkeley, the University of Illinois
and the University of Maryland with support from the National Science
Foundation.} \citep[][]{BIMA96}.  Phase calibration was done with
0224+671 and the ultracompact \ion{H}{2} region W3(OH) was used for
flux calibration.

Each OVRO track includes at least two phase centers separated by less
than the FWHM power points of the primary beams of the dishes (Table
\ref{Tobs}).  The pointings were naturally weighted and mosaiced using
the MIRIAD.  Quoted noise levels are the rms from line-free channels
of the spectral cube half-way between the map centers and edges.  The
true noise level is slightly lower ($\sim 10\%$) at the phase centers
and somewhat higher towards the edge of the primary beams due to the
mosaicing.  Subsequent data analysis was done with the NRAO AIPS.

Since interferometers act as spatial filters, it is important that
emission maps being compared have similar $(u,v)$ coverage, including
minimum baselines.  For the 3 mm lines observed with OVRO, the $(u,v)$
coverage is very similar. For the J=2--1 lines at $\lambda=1$ mm, the
$(u,v)$ coverage is consistent (C$^{18}$O(1--0) and C$^{18}$O(2--1)
were observed simultaneously), but scaled up by a factor of two from
their 3 mm counterparts ( $(u,v) = B_{proj}/\lambda$, projected
baselines in the east-west and north-south directions respectively).
The 2--1 transitions were tapered to match the $(u,v)$ range of the
corresponding 1--0 transition.  For the OVRO datasets, the minimum
(1--0) [(2--1)] baselines are $(u,v)_{min}\simeq 5.5~ k\lambda$ [$11~
k\lambda$].  Thus the images are insensitive to emission on spatial
scales $\gsim 35$--$40^{''}$ ($\sim$500--600 pc) for 110 GHz and
$\gsim 20^{''}$ ($\sim$300 pc) for 220 GHz.  For the $^{12}$CO(1--0)
transition, observed at BIMA, uv coverage is similar to the OVRO 3
mm uv coverage ($(u,v)_{min}~\sim 5.4~ k\lambda$).

We have estimated the amount of flux resolved out of each map due to
missing short spacings.  Single-dish spectra of $^{12}$CO(1--0),
$^{13}$CO(1--0), $^{13}$CO(2--1) and HCN(1--0) exist for Maffei 2 in
the literature (Table \ref{Tobs}.)  The interferometer map for each
line was convolved to the beamsize of the single-dish and then sampled
at the same pointing center.  Within uncertainties, all of the flux of
the CO(1--0) and $^{13}$CO(1--0) lines is detected.  The
$^{13}$CO(2--1) and HCN(1--0) maps recover $\sim$ 60\% of the single
dish flux, although the single-dish $^{13}$CO(2--1) flux measurement
is rather uncertain \citep[][]{WHEGGJRS92}.  We expect the
interferometer to recover fractions of C$^{18}$O flux similar to the
corresponding $^{13}$CO lines.

To generate integrated intensity maps, a mask was made by convolving
the channel maps to 10$^{''}$ resolution, then blanking regions of
emission $<$ 2$\sigma$.  This mask was then used to blank out
non-signal portions for the full resolution channel cube.  Velocities
from -160 km s$^{-1}$ to 100 km s$^{-1}$ were integrated, including
emission $>$1.2$\sigma$ in the full resolution channel maps.  For the
line ratio maps, the integrated intensity maps were convolved with an
elliptical Gaussian to the beam size of the lowest resolution maps
($^{13}$CO(1--0) and C$^{18}$O(1--0)).  This gives a resolution of
$\sim3.9^{''}$ for the line ratio maps.  Regions of emission $<$
3$\sigma$ in either line (2$\sigma$ for the
C$^{18}$O(2--1)/C$^{18}$O(1--0) map) were blanked in making the ratio
maps.  Because of the uncertainties in fluxes and absolute positions,
the ratio maps are estimated to be accurate to $\sim$ 35--40\% in
magnitude, and $\sim 2^{''}$ in position (excluding possible
systematic errors associated with differences in resolved-out flux).

Since Maffei 2 is within 1\degr\ of the Galactic Plane, and at
essentially zero redshift, we have to consider the possibility of
contamination from Galactic CO.  Galactic HI emission significantly
affects VLA images of Maffei 2 in the velocity range -70 km
s$^{-1}$--0 km s$^{-1}$ \citep[][]{HTH96}, with the strongest
absorption at -40 km s$^{-1}$. Galactic HI emission is more widespread
than CO emission, however, and what CO emission there is near Maffei 2
has very narrow lines ($<$ few km s$^{-1}$, based on our spectra from
the late NRAO 12 Meter Telescope.)  Inspection of our channel maps
reveals no obvious evidence of Galactic CO emission, and so we
disregard it.

\section{Molecular Gas in the Center of Maffei 2: Overview}
\subsection{Gas Morphology \label{overall}}

\begin{figure}
\vskip -3in
\epsscale{1.1}
\plotone{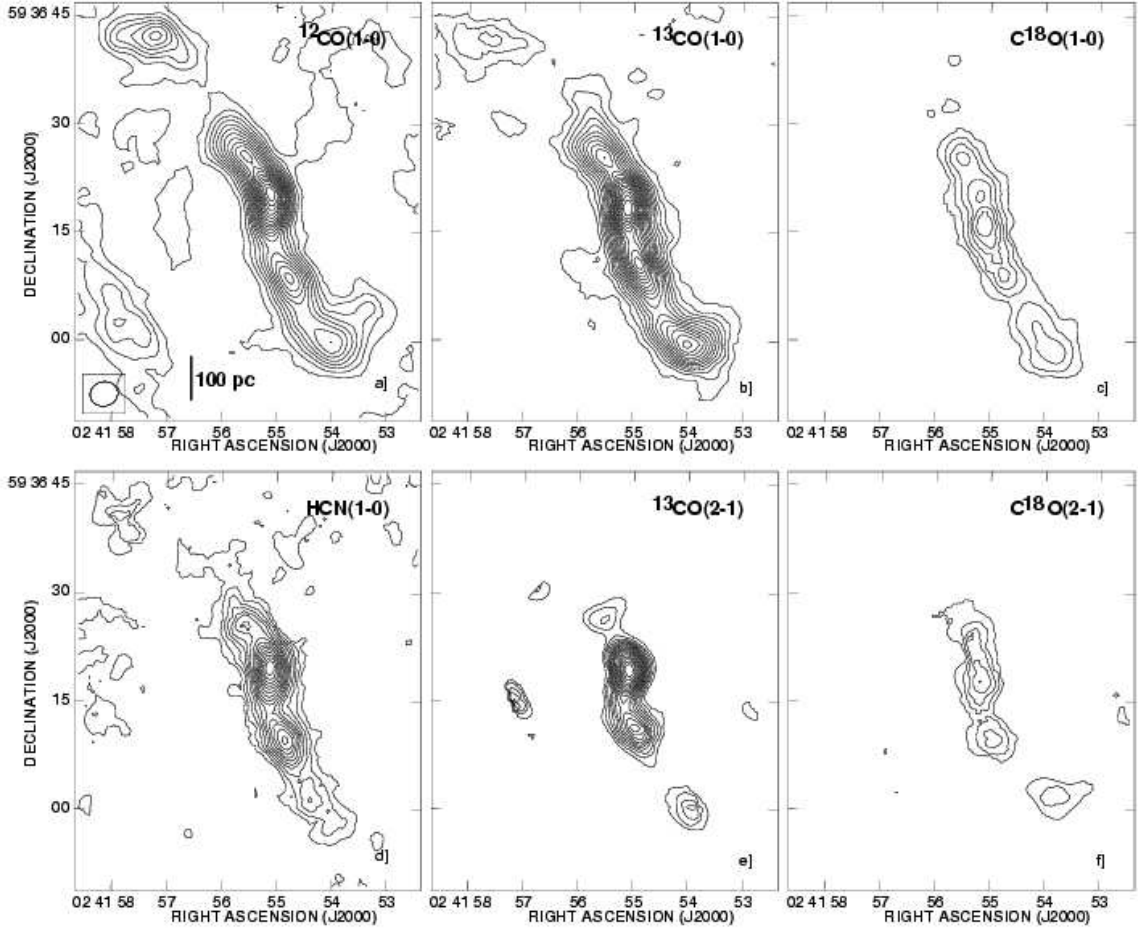}
\caption{Integrated intensity of CO isotopologues in Maffei 2. {\it
a)} CO(1-0), with contour levels of 10.0 Jy/beam km s$^{-1}$ (70 K km
s$^{-1}$, $\sim 3\sigma$).  {\it b)} $^{13}$CO(1-0), with contour
levels of 0.70 Jy/beam km s$^{-1}$ (5.4 K km s$^{-1}$, $\sim
3\sigma$).  {\it c)} C$^{18}$O(1-0), with contour levels of 0.70
Jy/beam km s$^{-1}$ (5.4 K km s$^{-1}$, $\sim 3\sigma$).  {\it d)}
HCN(1-0) with contour levels of 0.70 Jy/beam km s$^{-1}$ (8.3 K km
s$^{-1}$, $\sim 2\sigma$).  {\it e)} $^{13}$CO(2-1) with contour
levels of 2.5 Jy/beam km s$^{-1}$ (4.8 K km s$^{-1}$, $\sim 3\sigma$).
{\it f)} C$^{18}$O(2-1) with contour levels of 2.5 Jy/beam km s$^{-1}$
(4.9 K km s$^{-1}$, $\sim 2\sigma$).  All transitions have been
convolved to the resolution of the $^{13}$CO(1-0) transitions
($3.^{''}9\times 3.^{''}4$), shown in panel {\it a)}. }
\label{Finti}
\end{figure}

CO emission in Maffei 2 takes the form of two prominent and highly
inclined arms \citep[][]{I89}, which form the molecular bar, as shown
in the integrated intensity maps of Figure \ref{Finti}.  The emission
extends roughly 1\farcm5 (1440 kpc) along the major axis (Table
\ref{Tint}).  The brightest CO emission emerges from clouds within the
central 15\arcsec\ ($\sim$240 pc) of the galaxy.

\begin{figure}
\vskip -1.15in
\hskip .3in
\epsscale{0.9}
\plotone{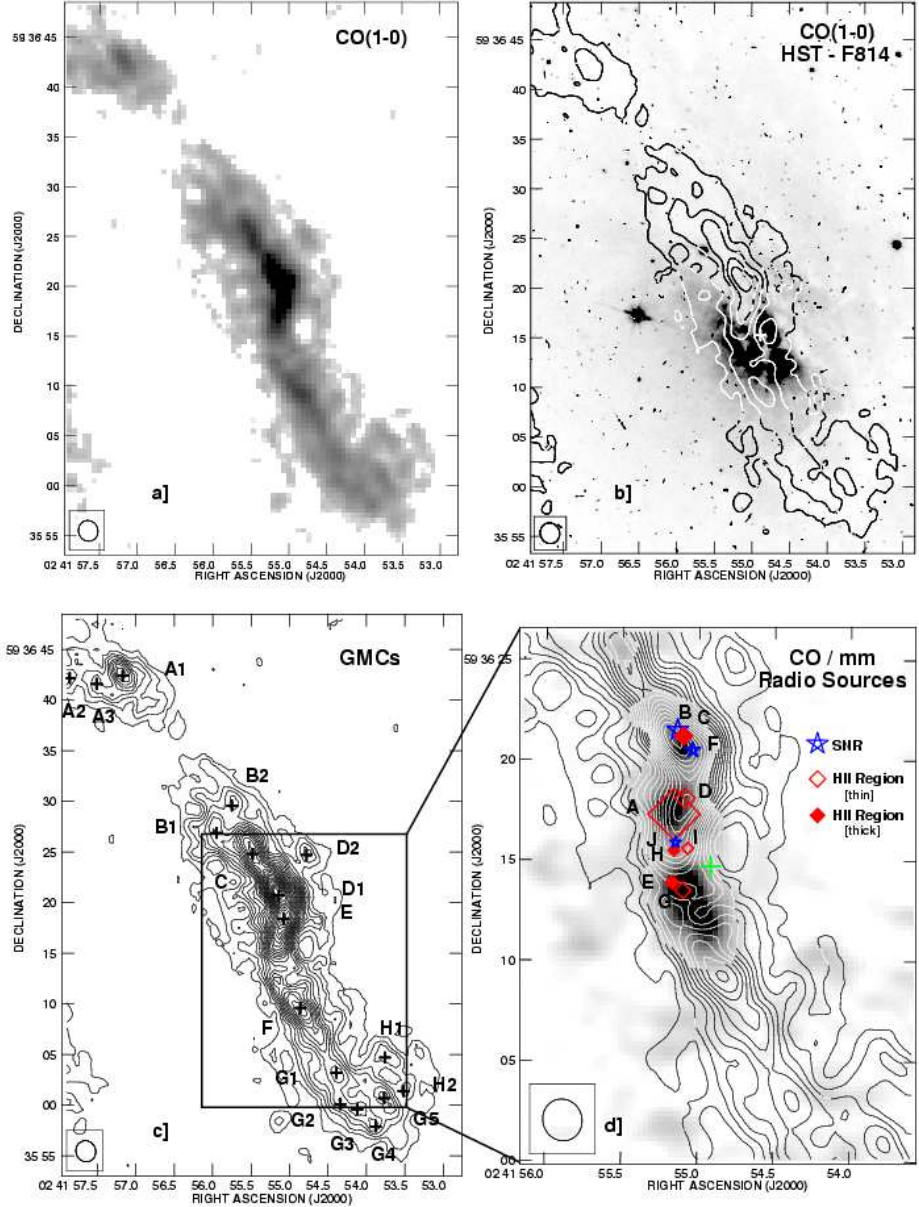}
\caption{{\it a)} Uniformly weighted integrated intensity maps of
CO(1-0) in Maffei 2.  The grayscale is in square root stretch ranging
from 4.0 Jy bm$^{-1}$ km s$^{-1}$ (95 K km s$^{-1}$) to 60 Jy
bm$^{-1}$ km s$^{-1}$ (1400 K km s$^{-1}$), for a $2.^{''}1\times
1.^{''}9$ beam.  {\it b)} The CO(1--0) map overlaid on the HST F814W
image of the nucleus.  Only contours 118 K km s$^{-1}$ $\times (1, 4,
8, 12, 16)$ are shown.  The cross marks the location of the dynamical
center (Table \ref{Tmaf}).  {\it c)} The CO(1--0) map with locations
of the fitted GMCs (Table \ref{Tgmc}) labeled.  Contours are the same
as {\it a)}.  {\it d)} A zoom in on the central region of the
molecular bar.  CO(1-0) is in contours (steps of 118 K km s$^{-1}$)
overlaid on the 89 GHz continuum image (grayscale).  Radio continuum
(2 cm) source identifications of \citet[][]{TTBCHM06} are labeled with
the symbols given in the legend. Symbol sizes are proportional to the
6 cm peak flux \citep[1 mJy bm$^{-1}$ = 1$^{''}$;][]{TTBCHM06}.  The
green cross marks the location of the dynamical center.}
\label{Fgmc}
\end{figure}

A higher (2\arcsec; $\sim$30 pc) resolution, uniformly-weighted image
of CO(1--0) (Figure \ref{Fgmc}a) shows that the bright CO peaks
resolve into GMCs.  The central two CO peaks become a nuclear ring of
radius $\sim$5\arcsec, or $\sim$80 pc about the dynamical center. The
eastern side of this ring is brighter in CO(1--0) but $^{13}$CO(1--0)
remains rather uniform.  The molecular arms are roughly linear
features running northeast and southwest, terminating at the central
ring.  Peak observed brightness temperatures, T$_{mb},$ reach $
\simeq$ 31~K, and are typically $\gsim$ 10~K across much of the arms.

Cloud properties---size, linewidth, temperature, mass---are derived
for the brightest molecular clouds in Maffei 2, using the
uniformly-weighted CO(1--0) image (Figure \ref{Fgmc}c).  Following
\citet[][]{MT01,MT04} (Table \ref{Tgmc}), a molecular cloud is defined
as a region of spatially and spectrally localized emission greater
than 2$\sigma$ in two adjacent channels, but need not necessarily be a
gravitationally bound entity.  Each cloud was fit from channel maps
that include only the gas over its localized velocity range.  Clumps
separated by one beamwidth or less are considered the same GMC.  Cloud
complexes are labeled A--H based on their locations in the lower
resolution maps; sub-clumps resolved in the higher resolution image
are numbered.

Most of the GMCs are resolved along at least one axis, with sizes of
$\sim$40--110 pc (Table \ref{Tgmc}). The clouds are significantly
elongated, with axial ratios often greater than 2.  Typical position
angles of the clouds, 20-60\degr, are very similar to the sky plane
position angle of the bar, $\sim$40\degr.  This is not an artifact of
the beam shape, since the beam is elongated perpendicular to the
bar. Nor is the elongation due to an underlying smooth gas component
along the bar, since fits come only from maps localized in velocity.
If the elongation of the clouds is a foreshortening effect due to the
high inclination of Maffei 2 (67\degr ; Table \ref{Tmaf}), then the
GMCs must be flattened perpendicular to the plane of the galaxy, that
is, disk-like, rather than spherical.  However, since similar
elongations are observed in the molecular clouds along the bar in the
nucleus of the face-on galaxy IC 342 \citep[][]{MT01}, we consider
cloud elongation along the bar more likely.  {\it The shapes of the
nuclear GMCs are clearly affected by their location within the bar.}

With the exception of two small GMCs (D2 and H1), cloud linewidths are
$> 50$ km s$^{-1}$ FWHM, and approach $\sim$ 100 km s$^{-1}$ in a
couple of locations.  If these clouds were in virial equilibrium, then
their individual masses would be in excess of $10^{7}
M_{\odot}$. However, these clouds are very unlikely to be in virial
equilibrium (\S \ref{gaskine}).  This is further demonstrated by the
fact that there is no correlation between the size $(\sqrt{ab})$ and
$\Delta v_{1/2}$ in Table \ref{Tgmc}.

The CO isotopologues generally follow the brighter CO emission, but
there are subtle differences.  Weak $^{13}$CO(1-0) emission
(Fig. \ref{Finti}b) extends to the map's edge, roughly along the major
axis of the large-scale near-infrared (NIR) \citep[][]{HMGT93a} and
the large-scale molecular bar \citep[][]{MW04}.  The $^{13}$CO(2-1)
emission (Fig.  \ref{Finti}e) is found mostly at cloud peaks due to
the higher critical density ($A_{ij}$) of the J=2--1 line, but some
diffuse gas not associated with the clumps is resolved out.  Little
emission off the trailing southwestern CO arm ridges (GMC H) is seen
in $^{13}$CO.

C$^{18}$O emission (Figs. \ref{Finti}c and \ref{Finti}f) follows
$^{13}$CO, but the C$^{18}$O linewidths are slightly narrower.  This
may be a critical density effect, with the lower opacity C$^{18}$O
more confined to the dense cores.  As a result of these spatial
differences, comparisons of C$^{18}$O line intensities with $^{12}$CO
and $^{13}$CO will slightly overestimate their true temperature
ratios.  Peak main-beam temperatures are T$_{mb}\simeq 0.5$ K
(T$_{mb}\simeq 1.0$ K) in a $3\farcs9\times 3\farcs4$ beam
($2\farcs4\times 2\farcs3$) for C$^{18}$O(1-0) (C$^{18}$O(2-1)).

The HCN(1--0) map is shown in Figure \ref{Finti}d.  Because of its
much larger electric dipole moment ($\sim$30 times CO), HCN has a
critical density nearly 1000 times higher than CO, and traces high
density gas.  HCN predominately traces the two inner peaks of GMCs D+E
and F.  HCN(1--0) emission falls off with distance from the center of
the galaxy faster than seen in any of the other lines (note
particularly GMC G).  There is also evidence that the HCN is more
strongly confined to the GMCs than CO or $^{13}$CO.  Apparently the
densest molecular gas is localized more strongly to the very center of
the galaxy.

\subsection{Star Formation in Maffei 2: Millimeter Continuum Images 
of \ion{H}{2} Regions and Dust \label{sfr}}

\begin{figure}
\vskip 0.6in
\epsscale{0.85}
\plotone{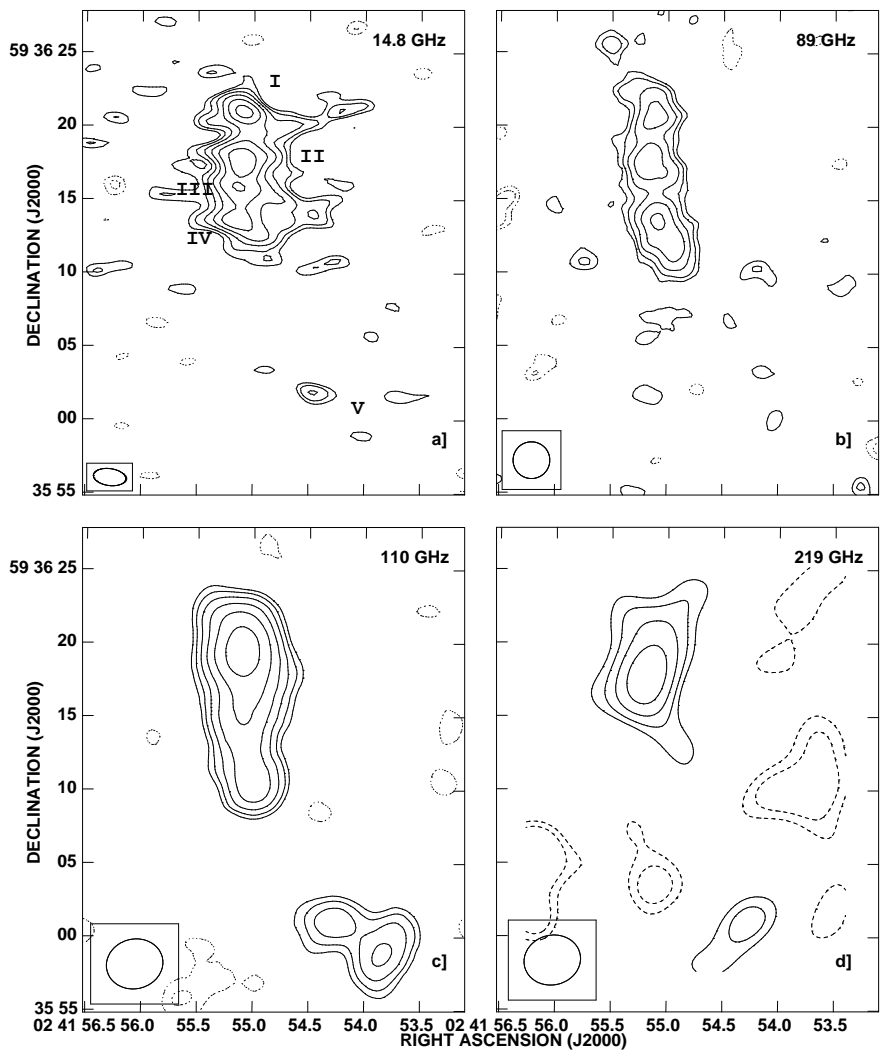}
\vskip -0.6in
\caption{Radio and millimeter continuum maps of the nucleus of Maffei
  2. Contours are $\pm 2^{n/2}$, n=0,1, 2... times the given lowest
  contour level.  Dashed lines are negative contours. {\it a)} 2 cm
  map of \citet[][]{TH94}, with the lowest contour at 0.45 mJy
  beam$^{-1}$ (2$\sigma$).  Roman numerals identify the five major
  continuum sources.  {\it b)} The 89 GHz (3.4 mm) map contoured in
  steps of 1.2 mJy beam$^{-1}$ (2$\sigma$).  Line contamination from
  HCN(1-0) and HCO$^{+}$(1-0) has been removed.  {\it c)} The 110 GHz
  (2.7 mm) map contoured in steps of 1.0 mJy beam$^{-1}$ (2$\sigma$).
  Line contamination from $^{13}$CO(1-0) has been removed.  {\it d)}
  The 219 GHz (1.4 mm) map convolved to the resolution of the 110 GHz
  map.  The lowest contour is 6.0 mJy beam$^{-1}$ (2$\sigma$).
  Beamsizes are indicated at the bottom left of each map, and in Table
  \ref{Tobs}.}
\label{Fcont}
\end{figure}

Continuum maps of Maffei~2 at 3.4 mm, 2.7 mm and 1.4 mm are presented
in Figure \ref{Fcont}, along with the 2~cm VLA maps from
\citet[][]{TH94}.  The 3.4 mm continuum has been corrected for the
contribution of HCN and HCO$^{+}$ lines within the bandpass, and the
2.7 mm continuum map for the contribution of $^{13}$CO. The advantage
of imaging continuum at 3 mm is that this is the part of the spectrum
where the free-free emission component from \ion{H}{2} regions is at
its maximum relative to other sources of emission, such as nonthermal
synchrotron and thermal dust emission.

There are three main 3.4 mm continuum sources near the center of
Maffei 2, with weaker sources towards the southwestern bar end (GMC G)
(Table \ref{Tcont}).  Four central sources are found at 2 cm; source
III has a non-thermal spectral index between 6 cm and 2 cm
\citep[][]{TH94}, and is predictably absent in the millimeter maps.
Sources I \& II are coincident with GMCs D and E and each have fluxes
of $\sim$ 5.8~mJy. Source IV is just north of GMC F and has a flux of
$\sim$ 5.0~mJy.  The non-thermal source III is not associated with any
of the bright GMCs.  At higher resolution these continuum sources
resolve into a collection of SNR and \ion{H}{2} regions \citep[Figure
2d;][]{TTBCHM06}.

\begin{figure}
\vskip -7in
\hskip -1.70in
\epsscale{1.4}
\plotone{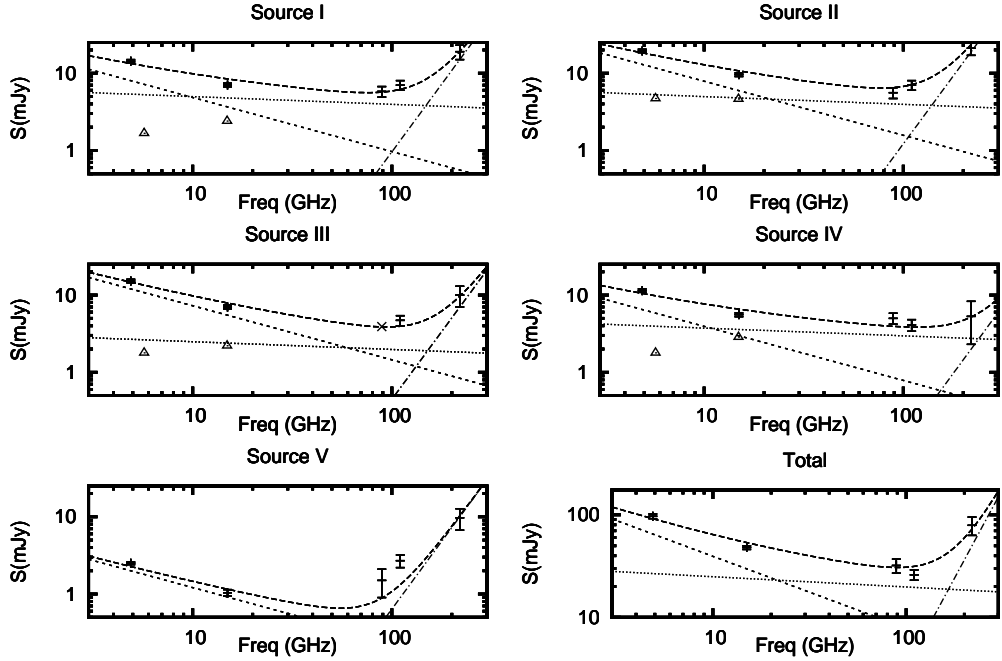}
\caption{Spectral energy distributions for the millimeter continuum
radio sources in Maffei 2.  Data are fit with synchrotron (dashed
line, $\alpha = -0.7$), free-free (dotted line, $\alpha = -0.1$) and
dust (dot-dashed line, $\alpha = +3.5$) emission components (see
text).  The cross in Source III represents an upper limit.  Triangles
mark the contribution of the compact emission  towards each source from 
higher resolution images \citep[][]{TTBCHM06}.}
\label{Fseds}
\end{figure}

Spectral energy distributions (SEDs) for each of the main radio
sources are shown in Figure \ref{Fseds}.  Three components are fit:
synchrotron, bremsstrahlung (free-free) and dust, with spectral
indices of -0.7, -0.1 and 3.5 ($\beta = 1.5$) respectively:
$$
S_{source}=S^{4.9}_{syn}\left( \frac{\nu}{4.885} \right)^{-0.7} +
 S^{89}_{ff}\left( \frac{\nu}{88.92} \right)^{-0.1} +
S^{219}_{d}\left( \frac{\nu}{219.3} \right)^{3.5}.
$$ Estimated fluxes for free-free and dust emission are recorded in
Table \ref{Tstarf}.  At cm wavelengths, the central continuum sources
have spectral indices between 6 cm and 2 cm, $\alpha^{6}_{2}$, of
-0.63 ($S_{\nu}\propto \nu^{\alpha}$), and therefore are dominated by
synchrotron emission.  The spectral index between 2 cm and 3.4 mm,
$\alpha^{2}_{3.4}$, flattens to -0.1 -- -0.3, as expected for
\ion{H}{2} regions dominated by free-free emission.  There can be
mixtures of synchrotron and free-free emission within our beam; we
have also shown cm-wave fluxes for the compact sources
\cite[][corrected for distance]{TTBCHM06} for comparison in Figure
\ref{Fseds}.  Our fits indicate that towards I, II and IV the 3 mm
continuum emission is dominated by the compact, free-free emission
sources.

The 1.4 mm continuum map is shown in Figure \ref{Fcont}d, convolved to
the resolution of the 2.7 mm map.  Emission peaks towards Source II at
$\sim$21 mJy beam$^{-1}$.  Continuum fluxes at 1.4 mm are larger than
the 3.4 or 2.7 mm fluxes, indicating a rising spectral index between
2.7 mm and 1.4 mm, $\alpha_{2.7}^{1.4}$, of +1.5.  The 1.4 mm emission
is therefore a mixture of free-free and dust emission, with the
predominance of dust varying with position.  The total flux associated
with dust emission, after removing the estimated thermal free-free
contribution, is $S_{1.4mm}\sim$7--19 mJy for each source.

\section{Gas Excitation and Opacity Across the Nucleus of Maffei 2}
\label{exotau}

Excitation temperatures are important for understanding molecular gas
properties and how they vary across the nucleus. The J=2--1 and
J=1--0 lines are sensitive to relatively cool gas in GMCs, and the low
J CO lines, especially CO(1--0), are thermalized in all but the lowest
density molecular clouds.  CO isotopologues thermalize at somewhat
higher densities ($\gsim 10^{3}$ cm$^{-3}$) due to their lower
opacity, making them excellent probes of gas excitation in this
density regime.

\subsection{Excitation Temperatures}

Excitation temperatures, T$_{ex}$, are constrained by the ratios of
integrated intensities of the J = 2--1 and J = 1--0 lines,
\begin{equation}
{\int ^{i}T_{21}dv \over \int ^{i}T_{10}dv}
\simeq{^{i}f_{21}(^{i}J_{21}(T_{ex}) - ^{i}J_{21}(T_{cmb}))
(1-e^{-^{i}\tau_{21}})\over ^{i}f_{10}(^{i}J_{10}(T_{ex}) - 
^{i}J_{10}(T_{cmb}))(1-e^{-^{i}\tau_{10}})},
\label{2110rat}
\end{equation}
where $^{i}J_{\nu}(T_{ex})~=~(h\nu/k)/(exp\{h\nu/kT_{ex}\}-1)$,
$^{i}\tau_{ul}$ and $^{i}f_{ul}$ are the optical depth and filling
factors of the $i$th isotopologue, respectively. We assume LTE
(constant T$_{ex}$ with J) throughout the cloud. Limitations of this
assumption are noted below.

The $^{13}$CO(2--1)/$^{13}$CO(1--0) line ratios for the nuclear bar
are shown in Figure \ref{Fratio} (Table \ref{Tratio}). Values range
from 0.3 to 0.8, corresponding to T$_{ex} \simeq 3$--6 K if $^{13}$CO
is optically thin, or up to 10 K, if completely thick.  This ratio
peaks towards the central two GMC complexes (GMC D+E \& F), and falls
with increasing radial distance.  C$^{18}$O(2--1)/C$^{18}$O(1--0) is
also higher at the central two emission peaks, with values of $\sim$
0.68--0.79.  C$^{18}$O is almost certainly optically thin.  From
C$^{18}$O(2--1)/C$^{18}$O(1--0), we obtain T$_{ex}\sim$5.4--6 K (Table
\ref{Tx_co}). There are several regions off the GMCs (between C and D;
H2) that have higher ratios.  The ratios are largest between clouds:
perhaps the intercloud gas is warmer than the GMCs (although emission
is weak here).  Figure \ref{Fpvrat} shows the average peak T$_{mb}$
ratios schematically as a function of velocity to show changes in
excitation along the central ring.  The eastern side of the central
ring has the highest T$_{ex}$ and this is the side closest to the
starburst.

T$_{ex}$ based on the isotopologues are significantly lower than both
the T$_{ex}$ implied by the single dish CO(2-1)/CO(1-0) line ratio of
$\gsim 2$ \citep[][]{SSMLP85} and the brightness temperatures,
T$_{mb}$, estimated from the high resolution CO(1--0).  Single-dish
CO(3-2)/CO(1-0) line ratios are also high, $\sim$1.3--1.8
\citep[][]{HTHM93b,DNTWW01}, as are $^{13}$CO(3-2)/$^{13}$CO(2-1)
ratios \citep[1.6;][]{WJBIMB93}.  These CO ratios suggest that there
is warm, optically thin gas with T$_{k} \gsim 50$ K.  Other molecules
indicate a range of gas temperatures.  From ammonia,
\citet[][]{TNKT00} find a rotational temperature of $T_{rot}\sim$30 K
that is constant across the field, and an ortho-to-para ratio
consistent with formation at 13 K.  \citet[][]{HMPFH00} obtain
$T_{rot}$= 85 K, based predominantly on the inclusion of the the high
energy metastable transition, (J,K) = (4,4).  However, they point out
that it is possible to fit the four lowest metastable transitions with
a cool component and a warm component.  \citet[][]{RHJM91} derived a
low T$_{ex}$ of 10 K from a multi-line study of HNCO.

Excitation of molecular clouds in the nucleus of Maffei 2 is complex,
and different molecular transitions will find different values for
T$_{ex}$, depending on where the molecules are found.  Some of the
differences in line ratios between CO, its optically thinner
isotopologues, and other molecular tracers, are due to the
isotopologues being subthermally excited relative to CO, such that
T$_{ex}$ $<$ T$_{k}$ because $n_{H_{2}} ~< ~n_{crit}(CO)$.  If the
densities determined from the LVG analysis are correct then the
T$_{ex}$ of the isotopologues imply kinetic temperatures of T$_{k}$ =
15 - 35 K (\S \ref{lvg}).  These are close to but still slightly
cooler than the (cool component of) ammonia.  Thus the bulk of
molecular gas, traced by the optically thinner species, is cool.
However CO emission is unlikely to be subthermal at these densities.
So the high single-dish CO line ratios are inconsistent with the
physical conditions of this component and require the presence of some
warmer gas.  Whether the emission from the high opacity CO transitions
originates in warmer envelopes of the clouds \citep[as in
IC~342,][]{THH93,MT01}, or from a compact, dense component (possibly
associated with the high temperature ammonia component) remains
unclear from the current data.  The low T$_{ex}$ of the higher
critical density HNCO may argue against the latter, but
spatially-dependent chemical effects may also be involved here.

\subsection{Opacity of the $^{13}$CO and C$^{18}$O Lines}

The opacity of the $^{13}$CO line is important for mass determinations
and the interpretation of brightness temperatures.  Based on the large
CO(3--2)/$^{13}$CO(1--0) ratio, \citet[][]{HTHM93b} estimated that the
$\tau_{13CO(1-0)} \sim 0.1$, or that $\tau_{CO(1-0)}$ is only 5--7, on
the assumption that CO(3--2) and $^{13}$CO(1--0) have the same
T$_{ex}$.  With higher spatial resolution this now appears not to be
the case.

Better opacity estimates are obtained by avoiding ratios taken between
lines with widely different opacities, particularly in situations
where temperature gradients and other non-LTE effects may be present.
The $^{13}$CO(1-0)/$^{18}$CO(1-0) integrated intensity ratio map of
Maffei~2 is shown in Figure \ref{Fratio}f.  Values range from
2.4--4.3.  The line ratio is lower than expected if the $^{13}$CO(1-0)
line (and the $^{18}$CO(1-0) line) have negligible opacities for
adopted abundance ratios of [H$_{2}$]/[$^{13}$CO]=$4.7\times 10^{5}$
and [H$_{2}$]/[C$^{18}$O]=$2.9\times 10^{6}$, or [$^{12}$CO]/[H$_{2}$]
= $8.5 \times 10^{-5}$, [$^{12}$CO]/[$^{13}$CO] = 60 and
[C$^{16}$O]/[C$^{18}$O]=250 \citep*{FLW82, HWLCM94, WR94,W99,
MSBZW05}. These isotopic abundances are typical of what is measured in
nearby starbursts \citep[eg.][]{HWLCM94}, and are within a factor of
$\sim$2 of the entire range observed in the Galaxy.  A
$^{13}$CO(1--0)/$^{18}$CO(1--0) line ratio of 3.0 implies
$\tau_{13CO(1-0)}\simeq 1$, for these isotopic abundance ratios.

Uncertainties in derived opacities depend sensitively on the true
[$^{13}$CO/C$^{18}$O] abundance ratio which may differ from the value
adopted here.  Both [CO/$^{13}$CO] and the [CO/C$^{18}$O] decrease
with stellar processing (assuming the CO isotopologues abundances are
proportional to their respective isotopic abundances).  Based on
Galactic disk studies, [$^{13}$CO/C$^{18}$O] is expected to decrease
from $\sim$7.5 in the local ISM to $\sim$4 in the inner kpc of the
disk, arguing for a decrease in the combined ratio with nuclear
synthesis \citep[eg.][]{W99,MSBZW05}.  Attempts to determine isotopic
abundances directly in starbursts obtain $^{13}$CO/C$^{18}$O $\simeq$
5, further supporting a lower ratio in high metallicity regimes
\citep[eg.][]{HWLCM94}.  On the other hand, Galactic Center (Sgr B2)
determinations arrive at anomalously high values of
[$^{13}$CO/C$^{18}$O] $\sim$10 \citep[eg.][]{LP90}.  Using the Galaxy
as a guide [$^{13}$CO/C$^{18}$O] should fall somewhere between 4 and
10.  We favor values on the low end of this range for the high
metallicity nucleus of Maffei 2 for three reasons: (1) The Galactic
disk gradient and the starburst values imply low ratios in heavily
processed locations. (2) The LVG models give consistent $^{13}$CO and
C$^{18}$O parameter space solutions for value [$^{13}$CO/C$^{18}$O]
$\simeq$ 4, but not for 10 (\S \ref{lvg}).  (3) There is marginally
significant evidence for lower CO(1--0)/$^{18}$CO(1--0) and
$^{13}$CO(1--0)/$^{18}$CO(1--0) line ratios along the central ring,
even towards the lower column density portions (Figure \ref{Fpvrat}).
This may represent direct evidence for enrichment of C$^{18}$O
relative to $^{13}$CO (and CO) in the immediate vicinity of the
nuclear starburst \citep[similar effects are seen in IC 342;][]{MT01}.

Modulo small differences in resolved-out flux or linewidth, we
conclude $\tau_{13CO(1-0)} \sim 1$ over the molecular peaks, but note
that systematic abundance uncertainties allow anything between $\tau
\ll$ 1 up to $\tau \sim$ 4. The fact that the CO(3--2)/$^{13}$CO(1--0)
line ratios of \citet{HTHM93b} imply much lower opacities must then be
a result of CO(3--2) emission (and likely CO in general)
preferentially originating from higher excitation gas than do
$^{13}$CO(1--0) and $^{18}$CO(1--0).

With $\tau_{13CO(1-0)}$ independently constrained from the
$^{13}$CO/C$^{18}$O ratio, comparisons between T$_{ex}$ derived from
eq. (\ref{2110rat}) (corrected for resolved out flux) and the observed
$^{13}$CO(1--0) peak brightness temperature, $^{13}$T$_{mb}$, gives
constraints on the filling factor, $^{13}f$, via $^{13}f ~\sim$
[$^{13}$T$_{mb}/^{13}J_{10}(T_{ex})](1-e^{-\tau_{13CO(1-0)}})^{-1}$.
Towards the molecular peaks $^{13}f$ $\sim$ 0.33 is estimated.  Given
the potentially large uncertainty in the estimate of the average
$^{13}$CO opacity, $^{13}f$ should be considered only indicative.  If
$\tau_{13CO(1-0)} \gg$ 1 then $^{13}f$ could be as low as $\sim$0.15.
The relatively bright $^{13}$CO(1--0) emission and the fact that
$^{13}f~\le$ 1 requires that $\tau_{13CO(1-0)} \gsim$ 0.33.

In summary, the J= 2--1 to 1--0 line ratios of the lower opacity CO
isotopomers imply LTE excitation temperatures of T$_{ex}\sim$3-10
K. Brightnesses of the higher opacity J $>$ 1 transitions of CO,
suggest that they preferentially sample more limited volumes of warmer
gas.  The opacity of the $^{13}$CO(1--0) transition appears to
approaches unity over much of the nuclear peaks.

\subsection{HCN}

\begin{figure}
\vskip -2in
\hskip -0.5in
\epsscale{1.1}
\plotone{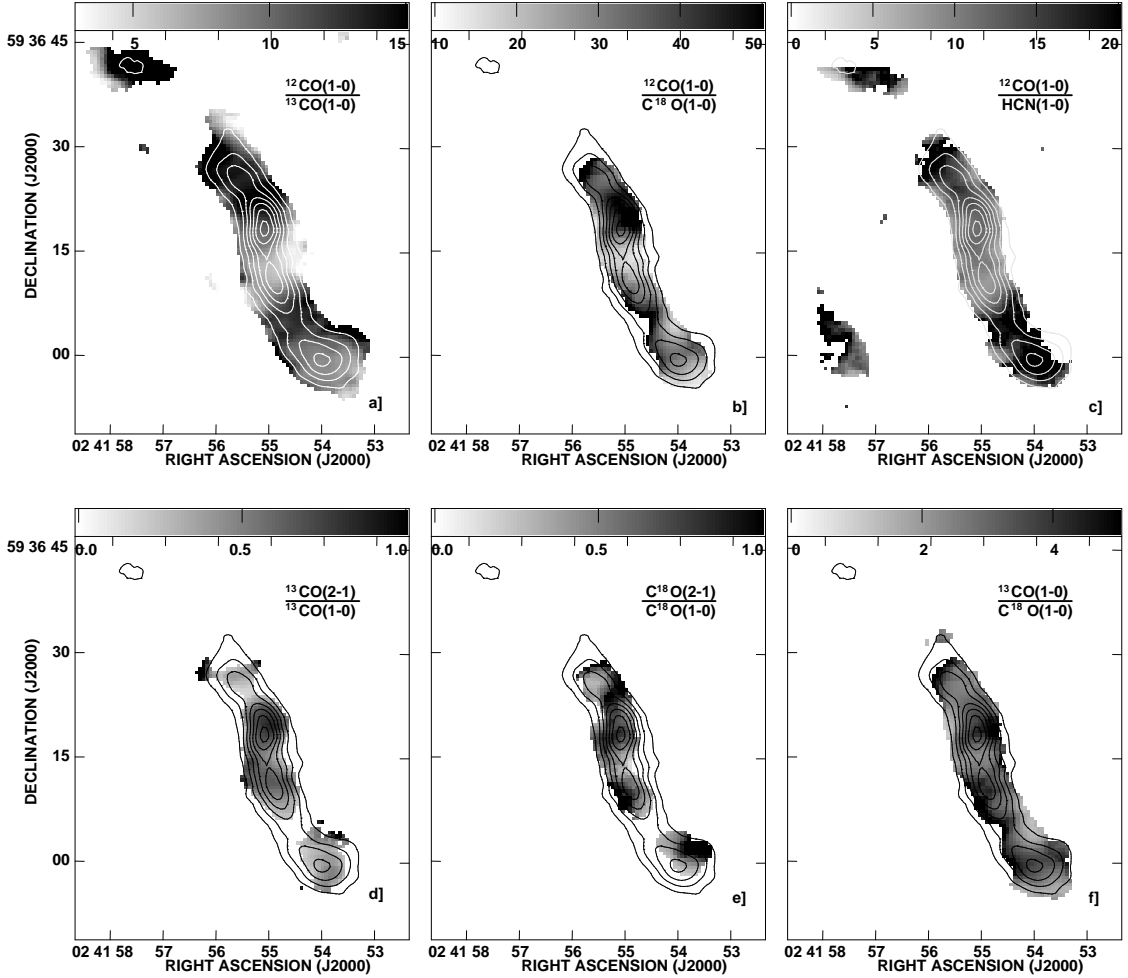}
\vskip -2.0in
\caption{CO and HCN line ratios in Maffei 2.  For comparison the
$^{13}$CO(1-0) integrated intensity is overlaid on all planes in
linear contours of 2.0 Jy/beam km s$^{-1}$ (15 K km s$^{-1}$).  The
resolution of all the ratio maps is $3.^{''}9\times 3.^{''}4$.  The
greyscale range of the ratio is noted in the wedge at the top of each
panel. (a) CO(1-0)/$^{13}$CO(1-0).  (b) CO(1-0)/C$^{18}$O(1-0).  (c)
CO(1-0)/HCN(1-0). (d) $^{13}$CO(2-1)/$^{13}$CO(1-0).  (e)
C$^{18}$O(2-1)/C$^{18}$O(1-0).  (f) $^{13}$CO(1-0)/C$^{18}$O(1-0).  }
\label{Fratio}
\end{figure}

\begin{figure}
\vskip -3in
\epsscale{1.0}
\plotone{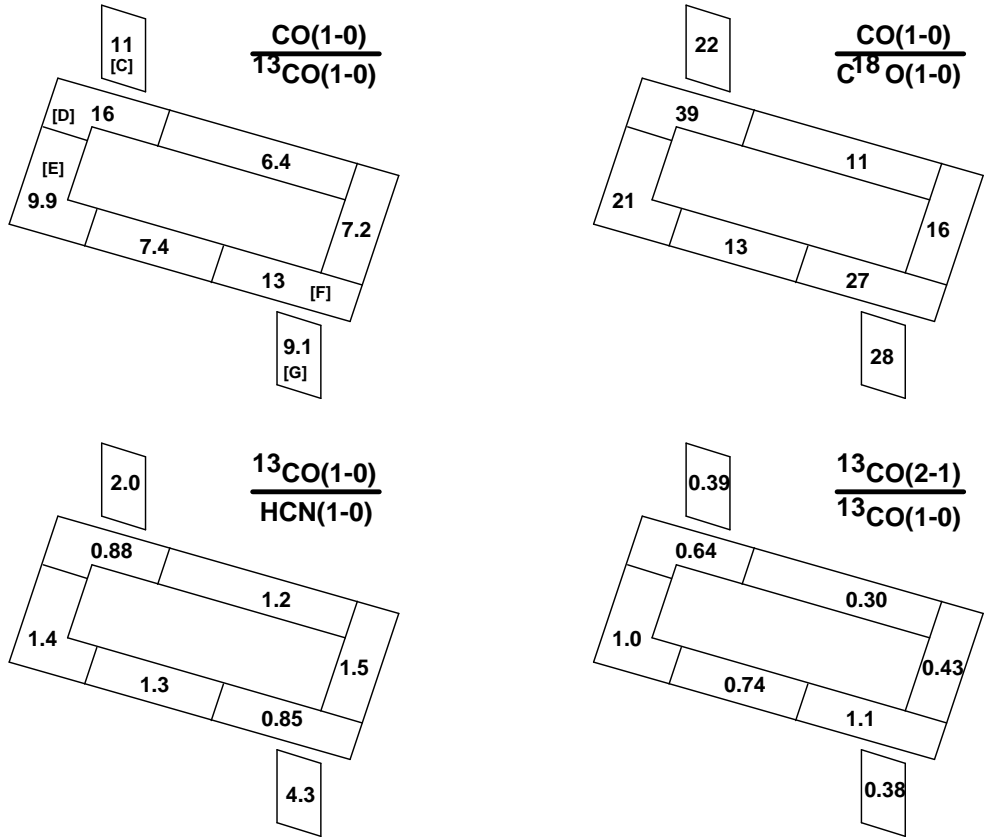}
\caption{Average peak T$_{mb}$ ratios for selected regions along the
central ring of the P-V diagram.  The schematic representation of the
P-V diagram has the same geometry as Figure \ref{Fpv}.  The letters in
parentheses in the upper left label the primary molecular peak in each
region.}
\label{Fpvrat}
\end{figure}

We also compare the distribution of dense gas traced by HCN to that of
the total gas traced by CO.  Figure \ref{Fratio}c, shows the
CO(1--0)/HCN(1--0) integrated intensity line ratio.  The ratio varies
from $\sim$8 at GMC F to $>$20 at the ends of the molecular bar.
Figure \ref{Fratio}c shows the general radial trend commonly seen in
galaxies, namely an increase in the CO/HCN intensity ratio as one
moves away from the star forming sites at the center
\citep[e.g.,][]{HB93,HB97,SNKN02}.  Sites of high
CO(1--0)/$^{13}$CO(1--0) ratios (particularly in the off-arm regions)
also have the highest CO(1--0)/HCN(1--0) ratios (Table \ref{Tratio}).
This provides evidence that the $^{13}$CO(1--0) is at least partially
sensitive to the density of the gas.

\section{Nuclear Gas Kinematics: The Parallelogram and a Bar Model 
\label{gaskine}}

Maffei 2 is a strongly barred galaxy with a disturbed morphology,
probably due to interaction with a nearby companion
\citep[][]{HMGT93a,HTH96,MW04}.  Position-Velocity (P-V) diagrams
based on lower resolution CO(1-0) data, show that the molecular gas in
the central regions reaches high ($\gsim$ 75 km s$^{-1}$) radial
velocities over very small projected radii \citep[][]{I89,HT91}.
\citet[][]{I89} has interpreted this feature as an expanding ring
generated by an explosive event some $\sim 5 \times 10^{6}$ yrs ago,
superimposed on a Keplerian component.  Since Maffei 2 is strongly
barred, it is reasonable to consider whether these motions are a
result of non-circular motions due to a barred potential.
\citet[][]{I89} argued against this based on the fact the position
angle of the ``molecular bar'' and the major axis of the galaxy are
close so that any non-circular motions from a bar would be in the
plane of the sky and therefore could not explain the motions observed.
However, if there are ILRs (so $x_{2}$ orbits exist) and / or a
nuclear bar rotated with respect to the large scale bar exists then
this is not the case.

\begin{figure}
\vskip -0.7in
\epsscale{1.0}
\plotone{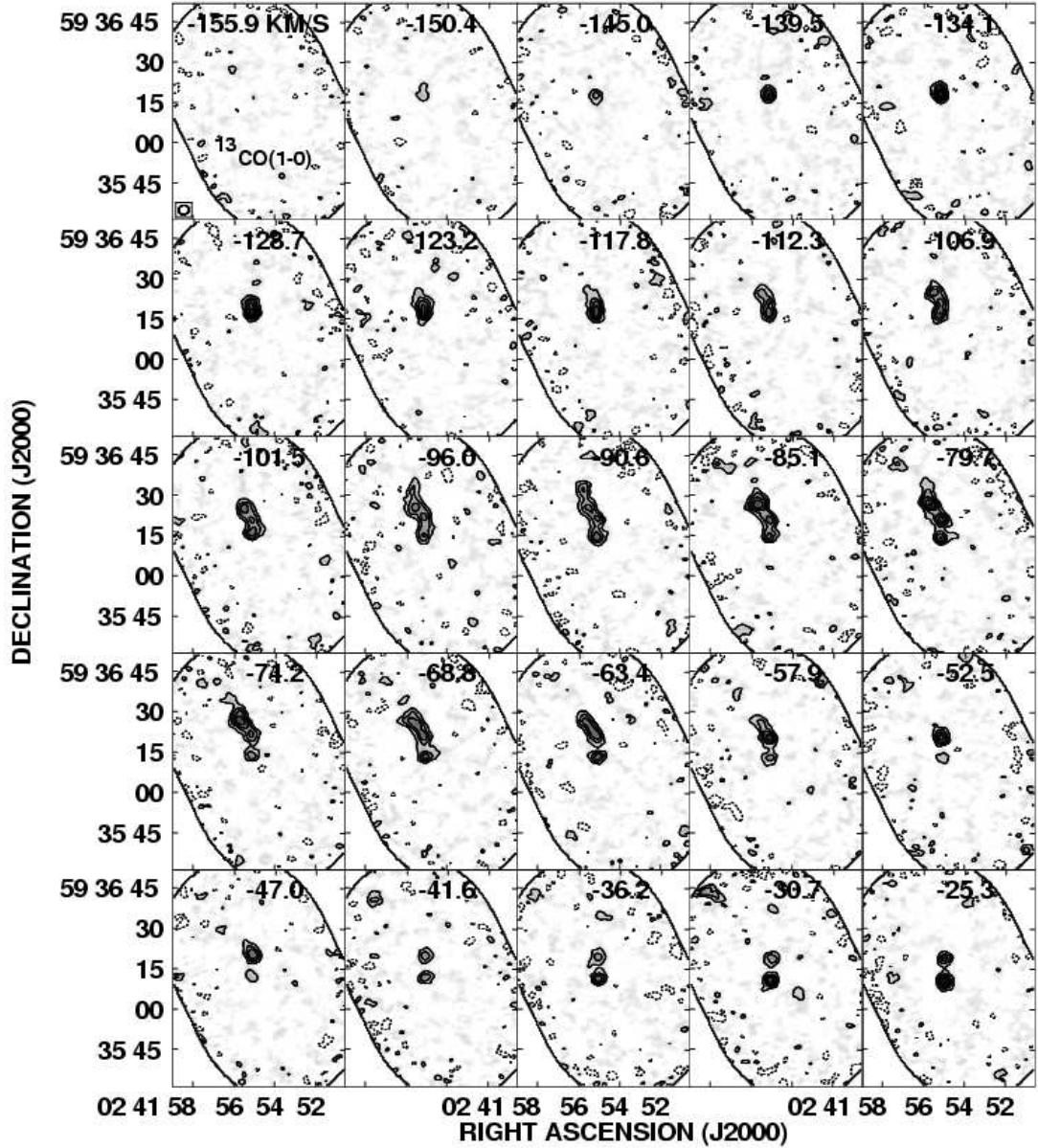}
\caption{The blueshifted half of Maffei 2's $^{13}$CO(1-0) channel
maps. Contours are multiples of  30 mJy/beam ( $2\sigma$). 
The beamsize is given in the bottom left of the first plane.}
\label{F13chana}
\end{figure}

\begin{figure}
\vskip -0.85in
\epsscale{1.0}
\plotone{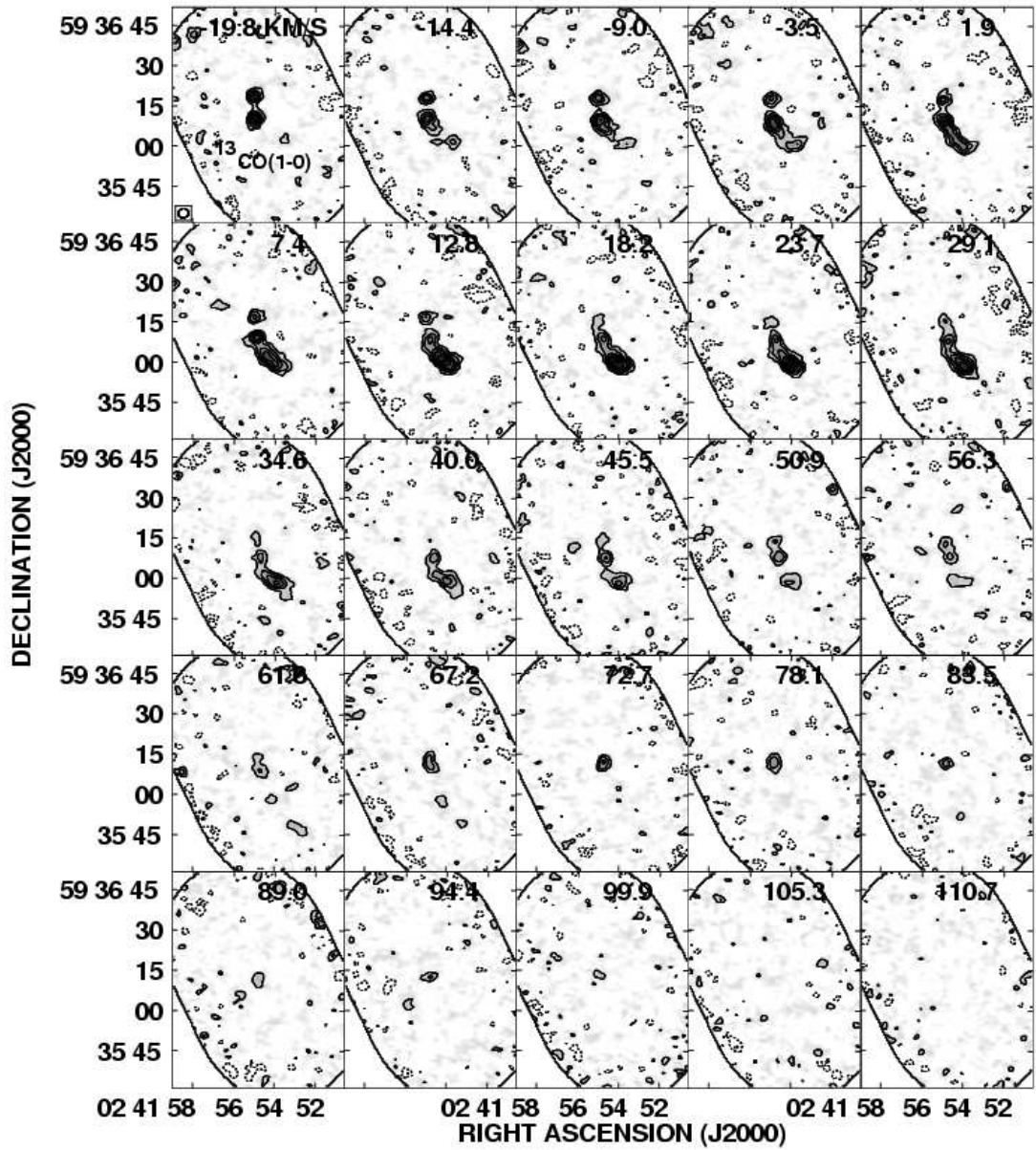}
\caption{The redshifted half of Maffei 2's $^{13}$CO(1-0) channel maps.
Contours given in Figure \ref{F13chana}.}
\label{F13chanb}
\end{figure}

\begin{figure}
\vskip -0.7in
\epsscale{1.0}
\plotone{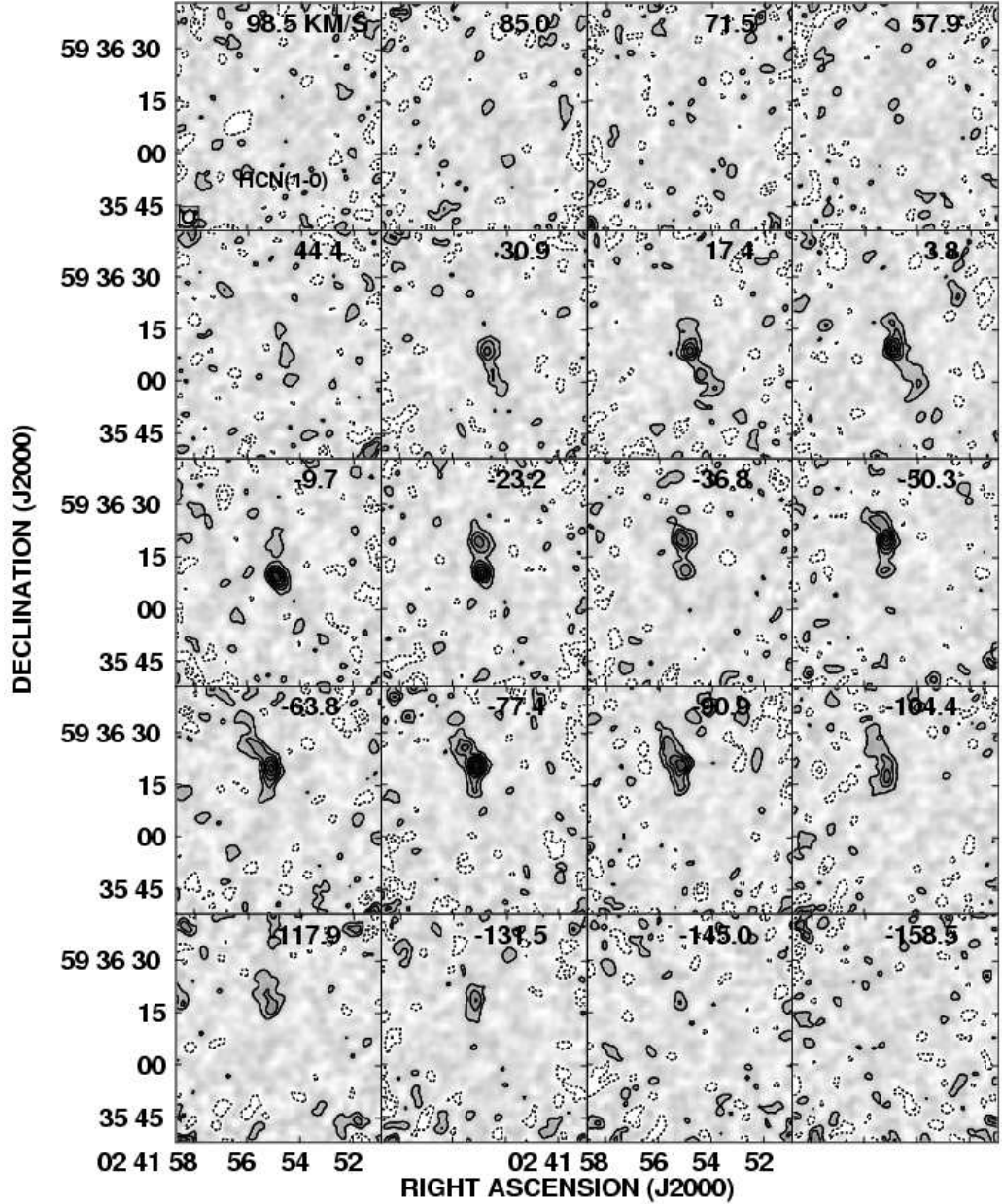}
\caption{Maffei 2 HCN(1-0) channel maps.  Contours are in increments
of the $2\sigma$ value, 20 mJy/beam.  The beamsize is given in the
bottom left of the first plane.}
\label{Fhcnchan}
\end{figure}

The gas kinematics can been seen in the channel maps of
$^{13}$CO(1--0) (Figures \ref{F13chana} and \ref{F13chanb}) and
HCN(1--0) (\ref{Fhcnchan}). $^{13}$CO(1--0) traces the velocity field
of the total column density, while HCN the velocity field of the dense
gas.  CO emission extends from $\rm v_{LSR}\sim -$160 to +100 km
s$^{-1}$ with blueshifted emission in the north. HCN is confined to
velocity ranges $\rm v_{LSR} \sim -$145 to +60 km s$^{-1}$.  If we
assume trailing spiral arms, the northern arm is the near arm,
consistent with the larger internal extinction there (Figure
\ref{Fgmc}b).

\begin{figure}
\vskip -3in
\epsscale{1.0}
\plotone{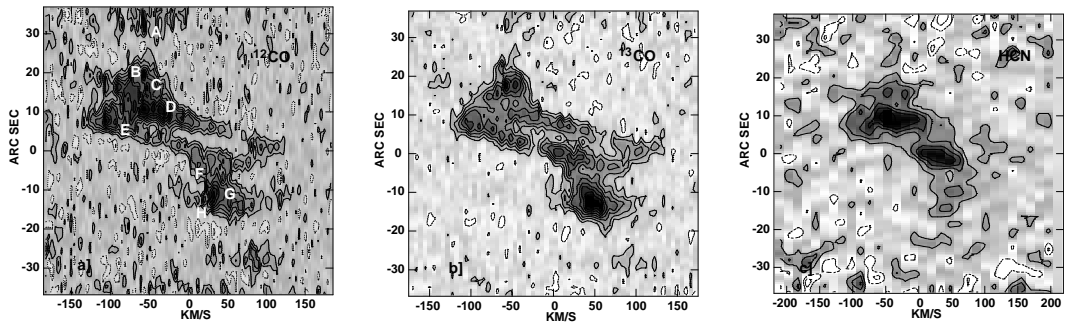}
\vskip -3in
\caption{The {\it a)}\ CO(1-0), {\it b)}\ $^{13}$CO(1-0) and {\it c)}\
HCN(1-0) Position-Velocity (P-V) diagrams taken along the major axis
of Maffei 2. The zero velocity corresponds to -20.5 km s$^{-1}$ (LSR).
The zero position corresponds to 02:41:55; 59:36:10.  Northeast along
the galaxy is at the top of the figure.  Contours are $\sim 2\sigma$.
Labels in {\it a)} identify the location of each GMC in position and
velocity.}
\label{Fpv}
\end{figure}

Position-velocity (P-V) diagrams for Maffei 2, shown in Figure
\ref{Fpv}, were made from the cubes by averaging the central 5$^{''}$
along the major axis of the galaxy ($p.a. = 206^{o}$; Table
\ref{Tmaf}).  We constructed major-axis P-V diagrams for CO(1-0),
$^{13}$CO(1--0) and HCN(1--0).  The CO(1--0) and $^{13}$CO(1--0) P-V
diagrams reveal an essentially complete ``parallelogram" associated
with the central ring.  The parallelogram has a width of
$\simeq$10\arcsec, a spatial extent of 20\arcsec (160 pc in radius),
and a total velocity extent of $\sim$250 km s$^{-1}$ (uncorrected for
inclination).  The $^{13}$CO-emitting gas along the central ring
appears to be nearly uniform.  CO(1--0), on the other hand, is
asymmetric with much brighter emission at the GMC D + E starburst
sites along the eastern side of the ring.  The cause of the asymmetry
is unclear, but since CO is so optically thick, its emissivity is more
susceptible to locally elevated kinetic temperatures or other non-LTE
surface effects. The uniformity of the $^{13}$CO is likely to be a
better indicator of gas surface density in this situation.  Beyond the
parallelogram, the velocity field is dominated by two peaks
corresponding to the ends of the molecular arms.

The velocity field of HCN(1-0) is significantly different from that of
CO.  The parallelogram is not apparent in the HCN P-V diagram, but
instead dominated by two peaks corresponding to the intersection of
the molecular arm emission and the parallelogram in the P-V diagrams.
Even along the central ring, HCN has a much lower covering fraction in
velocity space than CO.

\subsection{Bar Model for Maffei 2 \label{bar}} 

The new CO P-V diagrams (Figure \ref{Fpv}) have sufficient spatial
resolution to reveal that the ``oval''-shaped pattern is actually a
``parallelogram'' like that observed towards the Galactic Center
\citep[][]{BSWH88,BGSBU91}.  The similarity of molecular gas
kinematics in Maffei 2 to the Galactic Center, which is explained by
gas response to a barred potential \citep[eg.][]{BGSBU91,
HDMHWM98,RCMWA06}, leads us to construct a model of barred gas
response in Maffei 2.

We have modeled the gas distribution and kinematics in response to a
stellar bar using an analytic weak-bar model.  Such models are based
on treating the dissipational nature of gas with the addition of a
damping term proportional to the deviation from circular velocity
\citep[][]{W94,LL94,SOIS99}.  Despite their simplicity, these model
matches the structures seen in full hydrodynamical simulations with
surprisingly fidelity \citep[eg.,][]{LL94}.  The simplicity of an
analytic bar model permits us to quickly explore a wide range of bar
parameters.

\begin{figure}
\vskip -2in
\epsscale{1.1}
\plotone{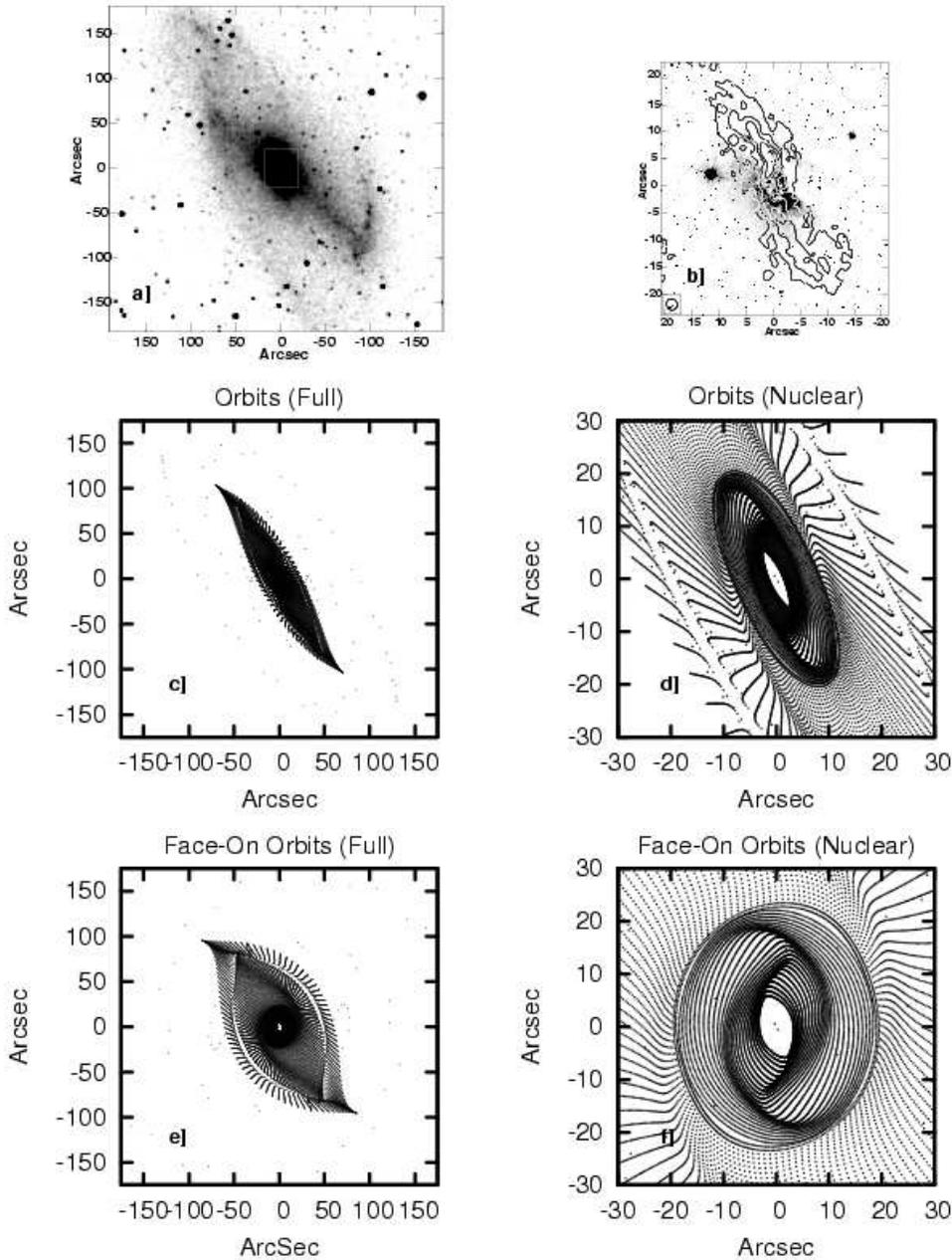}
\caption{An epicyclic double bar model for Maffei 2.  See \S \ref{bar}
for a description and Table \ref{Tbar} for parameters.  {\it a)} The
2MASS infrared K band image of Maffei 2.  {\it b)} Nuclear molecular
gas morphology traced by CO(1-0), showing only the 118 ($\times
1,4,8,12) $ K km s$^{-1}$ contours, overlaid off the 814W HST image.
{\it c)} A model of the inner portion of the large scale bar on the
same scale as {\it a)}.  Predicted positions of the gas peaks
follow the locations of highest point density.  Note that the
northeastern NIR arm is tidally disturbed, and as a result the model
agreement is poorer.  {\it d)} A zoom in on the nuclear bar portion
of the model.  {\it e)} As in {\it c)} except displayed face-on.  {\it
f)} As in {\it d)} except viewed face-on.}
\label{Fbarmorp}
\end{figure}

Our model is based on those of \citet[][]{W94} and \citet[][]{SOIS99},
with the following modifications. (1) The gas dissipation term is
extended to include azimuthal damping.  This is done by adding a term,
$2\mu \dot{\phi_{1}}$, to eq. (2) of \citet[][]{W94} analogous to
their eq. (1).  (2) The model is extended to include both a small
scale (hereafter ``nuclear") bar and a large scale bar, according to
the prescription of \citet[][]{MS97,MS00}.  (3) The axisymmetric
potential has been changed so that it generates a double Brandt
rotation curve coupled in size to the two bars.  The second component
of the potential is required both to match the flattening of the
rotation curve near the nucleus \citep[][]{HTH96} and to generate the
necessary existence of ILRs at $\sim$5\arcsec. The size and shape of
the large scale bar is set to match the NIR, CO and HI images of the
galaxy \citep[Figure \ref{Fbarmorp}a;][]{HMGT93a,MW04,HTH96}.  It is
assumed that the nuclear bar is co-planar with the large scale disk.

\begin{figure}
\vskip -6in
\epsscale{1.05}
\plotone{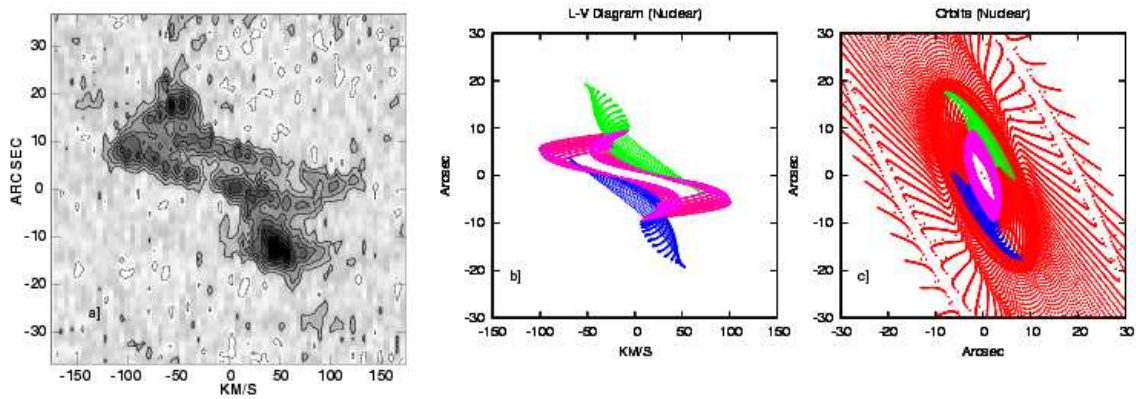}
\caption{The nuclear velocity field of Maffei 2.  {\it a)} The
$^{13}$CO(1-0) position-velocity diagram taken along the major axis 
(Figure \ref{Fpv}b).  {\it b)} The predicted P-V diagram for the
nuclear bar on the same scale as {\it a)}, showing the well known
'parallelogram' characteristic of barred motion.  Color coding
shows the portion of the velocity field associated with each of the
components labeled in {\it c)}.  {\it c)} Regions of high 
point density correspond to predicted (and observed) regions of gas 
concentration.  The three main regions of predicted gas concentration 
are color coded---the northern and southern arms are green and blue, 
respectively and the central ring is pink. }
\label{Fbarlv}
\end{figure}

\begin{figure}
\vskip -10in
\hskip -0.5in
\epsscale{1.5}
\plotone{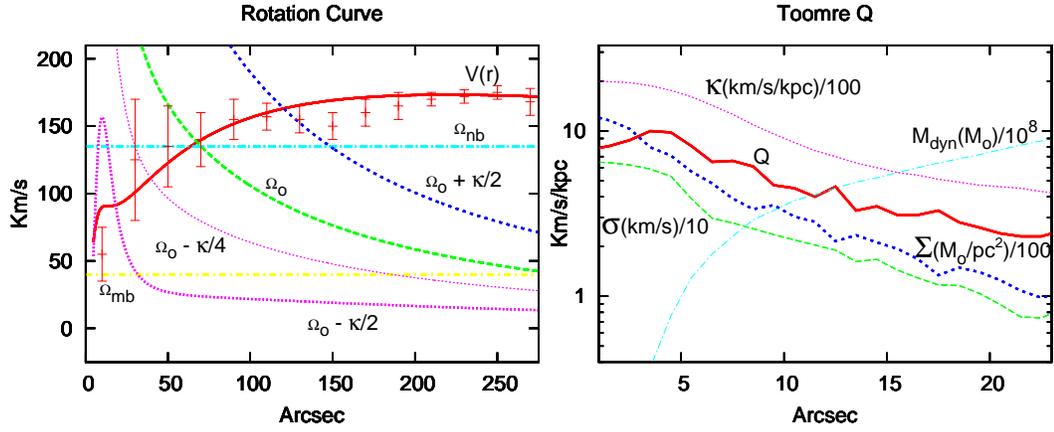}
\vskip -1in
\caption{Rotation curve of Maffei 2.  {\it a)} The model rotation
curve (thick red line) and observed HI rotation curve \citep[red
errorbars;][]{HTH96} is displayed.  Blue (yellow) dot-dashed lines
correspond to the pattern speeds of the nuclear (main) bar.
Associated resonance curves are labeled.  {\it b)} A plot of dynamical
properties as a function of galactocentric radius based on the modeled
rotation curve.  Shown are the azimuthally averaged observed velocity
dispersion (green), molecular gas surface density (dark blue) from
$^{13}$CO(1--0) assuming T$_{ex}$ = 5 K and an $\tau$ = 1 (see text),
epicyclic frequency (pink), implied enclosed dynamical mass (light
blue) and Toomre's Q (see text) (red).  Units for each curve is
given in the figure.}
\label{Fbarrot}
\end{figure}

Similarities between the model and both the true gas distribution
(Figure \ref{Fbarmorp}) and kinematics (Figure \ref{Fbarlv} \&
\ref{Fbarrot}) are excellent. Table \ref{Tbar} lists the fitted model
parameters.  In the weak bar scheme the molecular gas is predicted to
follow the sites of orbit crowding (higher density of dots in the
figure).  The model velocities match the observed pattern quite well.
In general, the models are robust to small changes in parameters as
long as two main requirements are met, (1) the combination of
potential/rotation curve and bar parameters are such that there are
two nuclear inner Lindblad resonances, oILR and iILR, and (2) the size
scale of each bar is close to the values chosen.  The first is
important because it is this condition that is required for
perpendicular $x_{2}$ orbits.  In our barred model, the perpendicular
$x_{2}$ orbits are vital to explain the large l-o-s velocities and
``parallelogram'' feature seen close to the center.  The second point
is important because it sets the scale of the features seen in the gas
(ie. the $x_{1}$ orbits run between the oILR and corotation).

Our model shows gas associated with the well known $x_{1}$ and inner
perpendicular $x_{2}$ orbits of barred potentials \citep[eg.][]{A92}.
A good fit for the nuclear morphology is achieved when the 4:1
ultraharmonic resonance of the nuclear bar is set equal to the oILR of
the large scale bar.  The pattern speed of the nuclear bar in Maffei 2
is thus much higher than the pattern speed of the larger bar, implying
that the nuclear bar is decoupled from the larger scale bar.  The
morphology is best matched with the nuclear bar rotated $\sim 10^{o}$
clockwise (as viewed from the perspective of Figure \ref{Fbarmorp}e)
relative to the large scale bar.

The ``parallelogram'' of the $^{13}$CO P-V diagram reveals additional
information about the gas associated with the closed nuclear bar
orbits.  The $^{13}$CO parallelogram is nearly complete.  This
suggests that, while not obvious in the integrated intensity map due
to the high inclination (though evidence is seen for it $\sim 5^{''}$
west of GMC E), the entire oval $x_{2}$ orbit region appears to
contain molecular gas.  The emission along the central ring in the P-V
diagram is nearly unresolved spatially, suggesting that only a very
small range of $x_{2}$ orbits are populated. Contrasts of $>$6 in
column density are seen between the molecular gas associated with
nuclear $x_{2}$ orbits and the very center of the galaxy, implying
that the vast majority of inflowing molecular gas is trapped at the
central ring and does not reach the core of the galaxy.  The
parallelogram suggests that a majority ($\sim$60 \%) of the $^{13}$CO
emission originates from gas residing on $x_{2}$ orbits with the rest
residing on the $x_{1}$ orbits.

The observed ``parallelogram'' is slightly wider along the position
axis than the model predicts.  This suggests that the inner $x_{2}$
orbits are slightly more circular than displayed in Figure
\ref{Fbarmorp}d.  This is likely due to the limits of the epicyclic
approximation at the very center of the potential.  Nearly circular
$x_{2}$ orbits are a common feature of the more complete
hydrodynamical simulations \citep[e.g.,][]{PST95}.

\section{Molecular Clouds in the Nuclear Environment: 
Cloud Properties as a Function of Location \label{lvg}} 

Armed with a reasonable kinematical model of the center of Maffei 2,
and having identified the sites of current star formation from mm
continuum maps, we can investigate the effects of environment on the
properties of molecular clouds. Densities, $n_{H_{2}},$ and kinetic
temperatures, T$_{k},$ of the individual nuclear GMCs were determined
by running Large Velocity Gradient (LVG) radiative transfer models
with the observed intensities and line ratios as inputs
\citep*[eg.][]{GK74,SS74,DCD75}.

We adopt single component LVG models for the clouds.  Three
independent parameters, $n_{H_{2}}$, T$_{k}$ and $X_{CO}$/$dv/dr$ are
varied over the ranges $n_{H_{2}}$ = $10^{1}$--$10^{6}~cm^{-3}$,
T$_{k}$ = 1.5--150 K, and $X_{CO}$/$dv/dr$ = 10$^{-3}-10^{-7.7}$
\citep[collision coefficients are from][]{DCD75}.  For each location,
seven (eight if including HCN) distinct measurements, the two isotopic
line ratios, the two $\Delta$J line ratios, the peak T$_{mb}$ of the
uniformly weighted CO(1-0) map and the ratio of cloud linewidth to
cloud size are used to constrain the model parameters.  The $\pm
1\sigma$ ranges do not include systematic uncertainties associated
with changes in $X_{CO}$/$dv/dr$ or more general uncertainties related
to the validity of the LVG approximation itself.  Additional model
solutions with $X_{CO}$/$dv/dr$ varied by a factor of $\pm$0.3 dex
were determined (not shown).  Increasing the velocity gradient (or
correspondingly decreasing the abundance) results in an increase in
derived densities by $\sim$0.3 dex and a decrease in derived T$_K$ of
$\sim$ 10 K.  This variation is indicative of the sensitivity of the
physical conditions to changes in the abundance or velocity gradient
at the level permitted by their systematic uncertainties (\S
\ref{exotau}).

\begin{figure}
\vskip -2in
\epsscale{1.0}
\plotone{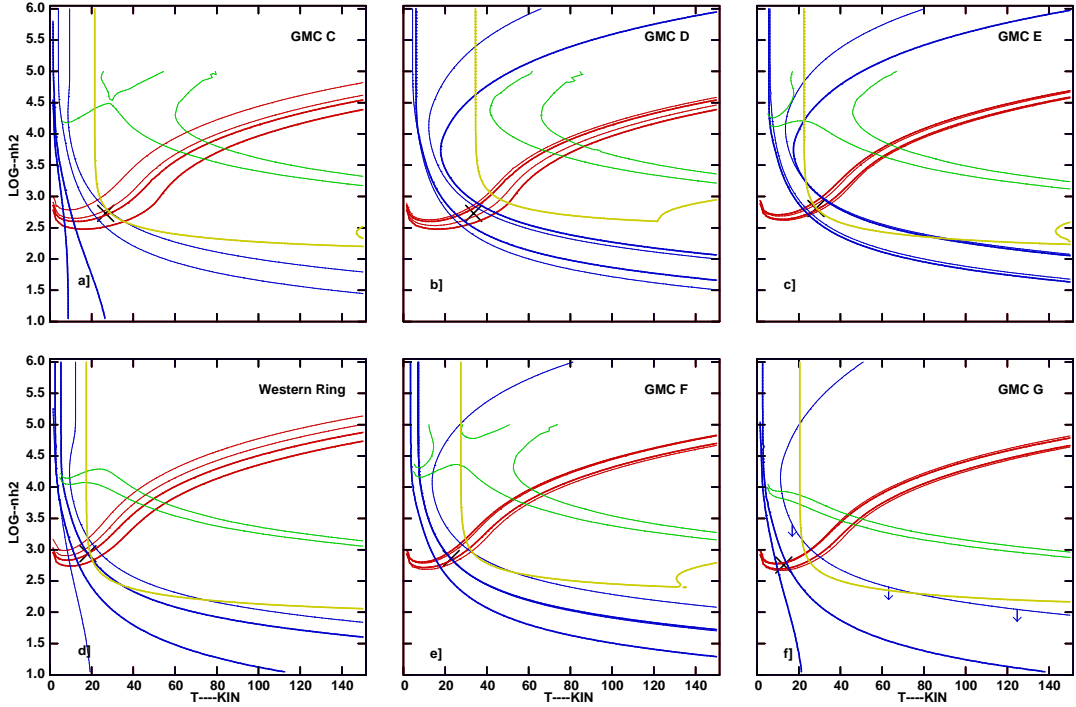}
\vskip -2.2in
\caption{LVG model solutions for six locations across the nucleus.
Abundances per velocity gradient, $X_{CO}/dv/dr$, between
$10^{-7.7}-10^{-3}$ have been modeled.  $^{i}X_{CO}\simeq$ $8\times
10^{-5}$ for [CO] / [H$_{2}$], (1/60)($8\times 10^{-5}$) for
[$^{13}$CO] / [H$_{2}$], (1/250) ($8\times 10^{-5}$) for [C$^{18}$O] /
[H$_{2}$], and a velocity gradient of 1 km s$^{-1}$ pc$^{-1}$ are
assumed (\S \ref{virial}).  Solutions are based on the $^{13}$CO(1-0)
resolution. Line ratios have been corrected for resolved-out flux.
Contours represent the 1$\sigma$ confidence solutions.  Red contours
display the isotopic line ratios (thick: CO(1--0)/$^{13}$CO(1--0);
thin: CO(1--0)/C$^{18}$O(1--0)) and blue contours the $\Delta$J line
ratios (thick: $^{13}$CO(2--1)/$^{13}$CO(1--0); thin:
C$^{18}$O(2--1)/C$^{18}$O(1--0)).  Green contours display the allowed
parameter space based on the $^{13}$CO(1--0)/HCN(1--0) line ratio,
assuming the filling factor of the HCN(1-0) emission is equal to that
of the $^{13}$CO and C$^{18}$O emission.  $X_{HCN}/dv/dr ~ = ~ 2.1
\times 10^{-8}$, is assumed, consistent with a Galactic Center HCN
abundance and a velocity gradient of $\sim$ 1 km s$^{-1}$ pc$^{-1}$
\citep[eg.,][]{PJBH98}.  The yellow line marks the CO(1--0) peak
brightness temperature assuming a filling factor of unity for the
uniformly weighted beamsize.  Crosses mark the adopted best fit
solutions.}
\label{Flvg}
\end{figure}

In Figure \ref{Flvg}, LVG model solutions for six locations across the
central bar are displayed.  Typical values of the velocity gradient
for the GMCs are $\sim$ 1--2 km s$^{-1}$ pc$^{-1}$ (Table \ref{Tgmc}).
We force abundance per velocity gradients of, $X_{CO}$/$dv/dr$ =
$10^{-6.12}$ ($10^{-6.75}$) for $^{13}$CO (C$^{18}$O), corresponding
to [CO/H$_{2}$] $\simeq 8 \times 10^{-5}$, [CO]/[$^{13}$CO] $\simeq$
60, [CO]/[C$^{18}$O] $\simeq$ 250 (\S \ref{exotau}), and $dv/dr~\sim $
1.5 km s$^{-1}$ pc$^{-1}$ (Table \ref{Tgmc}).  The $\pm 1\sigma$ range
for each ratio and the measured value for the CO(1-0) antenna
temperature constrain parameter space.

Average densities and kinetic temperatures implied by the LVG models
are $n_{H_{2}}$ $\simeq$ $10^{2.6-3.0}$ cm$^{-3}$ and T$_{k}$ $\simeq$
15--35 K.  All the mapped CO lines imply a consistent set of physical
conditions.  Densities derived from CO tend to be nearly constant
across the nuclear bar.  Clouds associated with the starbursts (D and
E) are slightly warmer than the others ($T_K\sim$ 30--40 K).  By
contrast, slightly cooler and denser values are derived for the
quiescent gas clouds on the nuclear $x_{2}$ orbit (`Western Ring';
Figure \ref{Flvg}) compared to the starbursting side of the central
ring (GMC E).  The solutions reproduce the observed brightness
temperature of the uniformly weighted CO(1--0), indicating a filling
factor of unity.  A unity filling factor for CO(1--0) is consistent
with that estimated from $^{13}$CO excitation given its larger beam
size.  Only GMCs D and F have predicted brightness temperatures
slightly {\it lower} than observed (by $\sim $50 \%), which could be
explained if the CO-emitting surfaces are somewhat warmer than the
bulk of the gas in these clouds.  The $^{13}$CO and C$^{18}$O
solutions are in excellent agreement, suggesting that the adopted
relative abundances are reasonable.

HCN LVG models were also run, for levels up to J=12 \citep[collision
coefficients from][]{GT74}.  Overlaid on Figure \ref{Flvg} are the
observed range ($\pm 1\sigma$) for the $^{13}$CO(1-0)/HCN(1-0) line
ratio, derived from the HCN(1-0) models.  An abundance per velocity
gradient, $X_{HCN}$/dv/dr = $2 \times 10^{-8}$ was assumed, consistent
with Galactic HCN abundances and $dv/dr$ = 1.5 km s$^{-1}$
\citep[eg.,][]{IGH87,PJBH98}.  A small correction for resolved out
flux has been made assuming emission is uniformly extended on scales
$\gsim 30^{''}$.  Densities derived from the HCN(1--0) line are about an
order of magnitude higher than fit from the CO isotopologues: HCN is
brighter than expected based on the CO-derived physical conditions.
This implies that (1) these molecular clouds have a significant
component of denser clumps from which the HCN originates, or (2) the
HCN abundance is much larger than has been adopted.  The absolute
abundance of HCN is not known well enough to eliminate the second
possibility, but the magnitude of the increase required ($\gsim
10\times$) makes it unlikely.  The HCN data suggests that while the
single-component LVG approximation yields internally consistent
solutions for the optically thinner isotopologues, which sample the
bulk of the molecular column density, it breaks down when including
the very high density gas.  The derived $n_{H_{2}}$ and T$_{k}$ should
then be treated as volume averages for the gas clouds traced in the
isotopologues, but not necessarily the whole of the ISM.  The
$^{13}$CO/HCN would then reflect the relative fraction of very dense
gas at each position \citep[e.g.,][]{KKV99,MT04}.

To summarize, molecular clouds in the central 300 pc of Maffei 2
averaged over $\sim$60 pc scales tend to be only modestly denser than
GMCs in the disk of the Galaxy.  How much denser depends on the exact
velocity gradient/abundance of CO and HCN present.  If the molecular
gas has a velocity gradient $\sim$ 1--2 km s$^{-1}$ pc$^{-1}$,
consistent temperatures and densities are obtained from the CO
emitting gas, at values of $<n_{H_{2}}> ~\simeq ~10^{2.75}~ cm^{-3}$
and T$_{k} ~\simeq ~ 15$ - 35 K for all GMCs.  At these densities, the
CO isotopologues are subthermally excited.  The HCN emission implies
subclumping.

\subsection{CO as a Tracer of Molecular Gas Mass:  The Conversion Factor 
in Maffei 2 \label{xco}}

There are indications that CO(1--0) is overluminous per unit mass of
$H_2$ gas in the nuclear regions of spiral galaxies relative to
Galactic GMCs, and thus use of the Galactic conversion $X_{CO}$ can
lead to overestimates of molecular gas masses in these systems.
\citep[e.g.][]{DHWM98,MT01,WNHK01,MT04}.  In this section, we compare
several different methods of estimating molecular gas column densities
to assess the validity of the conversion factor in the nucleus of
Maffei 2.

\subsubsection{Molecular Gas Column Densities from the Optically thin 
CO Isotopologues, Dust Continuum, and the Virial Theorem }
\label{virial}

Optically thin lines of CO isotopologues allow estimates of the
molecular gas column density directly by summing up the emission from
each molecule. These estimates depend only on the knowledge of
relative CO abundance and excitation.  For LTE
\citep[eg.][]{SSSCMLP86},
$$
N(H_{2})_{^{i}CO}=5.75\times 10^{17}~cm^{-2}~\frac{[T_{ex}+ 0.92]}{^{i}\nu_{G}^{2}}~
{[~H_{2}~] \over [^{i}CO]}~ {e^{^{i}T_{o} \over T_{ex}}}~ \times~~~
$$
\begin{equation}
~~~~~~~~~~~~~~~~~~~~~~~~~~~ \left( \frac{^{i}\tau}{1-e^{-^{i}\tau}} \right)
~ I_{^{i}CO}~(K~ km~ s^{-1}).
\end{equation}
where $[~H_{2}~] \over [^{i}CO]$ is the abundance of the isotopologue,
$^{i}\nu_{G}$, $^{i}\tau$ and $^{i}T_{o}$ are the frequency (in GHz),
the opacity and the characteristic temperature ($h\nu_{o}/k$) of the
particular transition ($^{i}T_{o} = 5.29$ for $^{13}$CO(1-0) and 5.27
for C$^{18}$O(1-0)).  We calculate T$_{ex}$ separately from the
$^{13}$CO(2-1)/$^{13}$CO(1-0) line ratio assuming that $^{13}\tau_{CO}
\simeq 1$ (\S \ref{exotau}), and the C$^{18}$O(2-1)/C$^{18}$O(1-0)
line ratio.  These values and the H$_2$ column densities derived from
them are presented in Table \ref{Tx_co} for each peak (T$_{ex}$ = 5 K
is assumed for positions beyond the (2-1) primary beam).  Under LTE,
T$_{ex}$ values turn out to be almost independent of opacity over the
range of 0 -- 5 (changing by $<$ 20 \%) when J = 2--1/J=1--0 line
ratios are around the observed value of $\sim 0.7$.  Therefore
systematic uncertainties in N$_{H_{2}}$ stem primarily from
uncertainties in the assumed isotopic abundances.  Abundances are
expected to be within a factor of $\lsim$2 of the adopted values (\S
\ref{exotau}).  

Values of N$_{H_{2}}$ range from $ <0.61~(1.3)$--$10~(6.5)\times
10^{22}$~cm$^{-2}$ (Table \ref{Tx_co}), based on $^{13}$CO (C$^{18}$O)
fluxes, with corresponding mass surface densities of $\Sigma ~ \simeq
~ <$130 (280)--2200 (1400) $M_{\odot}~$pc$^{-2}$.  Emission from the
fainter C$^{18}$O line does not extend to the lower column densities,
otherwise the predictions of column densities from the two species
agree to within a factor of 2.

Dust continuum emission has also been detected at $\lambda =$1.4 mm
towards the several of the GMCs, which gives another estimate of the
molecular gas mass. After accounting for the free-free contribution
(Table \ref{Tcont}), dust fluxes, $S_{1.4mm}$, range from 7--19 $\pm$3
mJy for each cloud.  Assuming a gas to dust ratio of 100 by mass, the
gas mass is related to the 1.4 mm dust continuum flux by
\citep[eg.,][]{H83}:
$$
M_{gas}(1.4~ mm)~=~306~ M_{\odot} \left(\frac{S_{1.4mm}}{mJy} \right)
\left(\frac{D}{Mpc} \right)^{2}
$$
\begin{equation}
~~~~~~~~~~~\left(\frac{\kappa_{\nu}}{cm^{2}~g^{-1}} \right)^{-1}
\left(e^{\frac{10.56}{T_{d}}}-1\right),
\label{dusteq}
\end{equation}
where $\kappa_{\nu}$ is the dust absorption coefficient at this
frequency, $S_{1.4mm}$ is the 1.4 mm dust flux, D is the distance and
T$_{d}$ is the dust temperature.  The dust opacity, $\kappa_{\nu}$, at
1.4 mm is taken to be $\rm 3.1 \times 10^{-3}~ cm^{2}~g^{-1}$,
uncertain by an estimated factor of four \citep{PHBSRF94}.  We adopt
T$_{d}=40$ K based on the FIR colors \citep{RH83}.  The dust
temperature applicable to the 1.4 mm observations could be lower than
this if a cool dust component undetectable shortward of 160$\mu$m
exists.  The existence of a cooler dust component would cause us to
underestimate the implied molecular gas mass. This is likely more
important away from the nuclear region. Dust masses for the cloud
peaks are listed in Table \ref{Tx_co}.

Virial masses can be derived from linewidths, following the treatment
of \citet{MT01} since the individual GMCs in the center of Maffei 2
are resolved.  Virial masses are given in Table \ref{Tx_co}. They have
an intrinsic uncertainty of about a factor of two due to internal
cloud structure.  In addition, if systematic motions such as cloud
streaming motions or non-circular bar motion, are present within a
single beam (almost certainly the case; \S \ref{gaskine}), the
linewidths due to the internal gravity will be overestimated. In
short, the virial masses will be upper limits to the true cloud
masses.

Finally, as a crude consistency check, a column density,
$^{L}N_{H_{2}}$, is calculated from the LVG model derived densities,
by averaging the number density over an assumed depth of
$\sqrt{\theta_{a}\theta_{b}}$.  These values are also recorded in
Table \ref{Tx_co}.  These values represent upper limits to N(H$_{2}$)
if $n_{H_{2}}$ is confined to a fraction of this volume.

\subsubsection{The CO Conversion Factor in Maffei 2}

We can compare the three different column density
estimates---optically thick CO ($^{Xco}N_{H_{2}}$), optically thin
$^{13}$CO ($^{13}N_{H_{2}}$) and C$^{18}$O ($^{18}N_{H_{2}}$), and
dust ($^{D}N_{H_{2}}$) --- to estimate a CO conversion factor,
$X_{CO},$ for the nucleus of Maffei 2.  Column densities based on the
CO isotopologues (Table \ref{Tx_co}) are lower than the those derived
from CO(1-0) intensities using the Galactic value of $X_{CO}$ $\simeq$
$2\times 10^{20}$ $\rm cm^{-2}$ $\rm (K~km~s^{-1})^{-1}$ \citep{Set88,
Het97,DHT01}.  Values from the thin C$^{18}$O lines are $\sim$2--4
times lower than the X$_{CO}$ estimates.  If we require that H$_{2}$
column densities derived from opacity-corrected $^{13}$CO(1--0) and
C$^{18}$O(1--0) agree (thereby constraining $\tau_{13CO(1-0)}$) then
Galactic values of the conversion factor can be reached only for
[CO/C$^{18}$O] $\gsim$ 600.  Given the high metallicity environment of
the nucleus of Maffei 2, this seems unlikely.  Away from the nucleus
there is some evidence for another factor of two further decrease in
the conversion factor; however, statistical uncertainties are at least
this large due to weak emission and T$_{ex}$ not being determined
towards these locations.

Uncertainties estimated for the gas column derived from dust emission
are higher than for the isotopologues, but they too tend to support
lower gas columns than predicted by the Galactic $X_{CO}$.  Dust-based
gas masses are also lower than the $X_{CO}$ values by factors of
$\sim$2--4 for the adopted dust parameters towards the detected GMCs
(Table \ref{Tx_co}).  Gas column densities estimated from the dust are
in good agreement with the opacity corrected $^{13}$CO estimates
except for GMC E, but trend a factor of $\sim$30 \% higher than those
from the C$^{18}$O isotopologues.  This is an indication that the
uncertainty in these column densities are at least this large.  While
these methods are different, we do not claim they are completely
independent, because there may be hidden correlations between, say, CO
relative abundance and dust to gas ratio. But the dependences on
metallicity and other factors such as temperature are not necessarily
the same for these mass tracers.  That the gas column densities
estimated from the dust and C$^{18}$O are both low provides additional
confidence for the assertion that the gas column densities are
overestimated by the Galactic value of $X_{CO}$.

Column densities obtained by averaging the virially derived masses are
higher than the other methods, which is not surprising.  Linewidths in
the central region (particularly GMCs C, D, E and F) include two
distinct components moving on completely different orbits (Figure
\ref{Fpv}), and so systematic motions within the $\sim$60 pc
(line-of-sight) beam due to the bar orbits (\S \ref{bar}) cause an
observed linewidth larger than random gravitational motions within the
clouds would imply. Because of the presence of this motion, we expect
that virial methods using observed linewidths to be severe
overestimates of the cloud masses.

{\em In summary, we conclude that the conversion between
$^{12}$CO(1-0) and H$_{2}$ column density applicable to the central
region of Maffei 2 is $^{M2}X_{CO}$ $\simeq$0.5 --1.0~$ \times
10^{20}$ $cm^{-2}~(K ~km~s^{-1})^{-1}$, $\sim$2--4 times lower than
the Galactic value, with uncertainties of $\sim$100 \%.}

\section{A Nuclear Bar-Driven Starburst in Maffei 2}

\subsection{Star Formation Rates and Efficiencies \label{SFR}}

The rotation curve from the double bar model can be used to estimate
dynamical masses directly without having to try to remove the
non-circular velocity component (Figure \ref{Fbarrot}b).  The
dynamical mass over the central ring is M$_{dyn}(R=7^{''})=2.1\times
10^{8}~M_{\odot}$.  The molecular mass estimated from C$^{18}$O over
the same region is 6.9$\times 10^{6}~M_{\odot}$.  Dynamical masses for
the central $20^{''}$ radius are M$_{dyn}(20^{''})=7.3\times
10^{8}~M_{\odot}$, while the molecular mass over this region is
2.1$\times10^{7}~M_{\odot}$.  Molecular mass fractions are thus
$\sim$3\% percent over much of the central molecular bar.  Molecular
mass fractions scale as the distance, so the uncertainty in the
distance to this galaxy (\S 2) can change these values by up to a
factor of 2.  Resonant structures, such as the molecular bar observed
in the nucleus, are probably driven by the stellar potential rather
than the gas.

Lyman continuum ionization rates, N$_{Lyc}$ \citep[for T$_{e}=10^{4}$
K; e.g.][]{MH67}, and star formation rates based on the 89 GHz
continuum are given in Table \ref{Tstarf}.  To produce the total
observed free-free emission across the central 30$^{''}$, the
excitation of 2600 effective O7 \citep[][]{VGS96} stars is required.
A significant fraction of this ionizing flux ($\sim
1.4\times10^{52}~s^{-1}$, or $\sim$ 1400 ``effective'' O7 stars)
arises near the two central molecular peaks (GMCs D1+E and F).
Towards radio continuum sources, I, II and IV, the local star
formation rates are 0.05, 0.05 and 0.04 M$_{\odot} ~yr^{-1}$,
respectively, based on the conversion between $N_{Lyc}$ and SFR of
\citet[][]{K98}.  These values match the star formation rate predicted
from the HCN(1--0) luminosity using the relationship that
\citet[][]{GS04} have derived from large scale HCN measurements.  The
relationship between HCN(1--0) luminosity and star formation on 60 pc
scales in Maffei 2 is the same as that observed on kpc scales in
luminous infrared galaxies.

The ionization rate across the nuclear bar corresponds to a massive
star formation rate of $\sim$ 0.26 M$_{\odot}$ yr$^{-1}$ with
$\sim$0.14 M$_{\odot}$ yr$^{-1}$ originating from the nuclear ring.
At this rate the molecular gas in the ring could sustain the current
star formation rate for $\sim 5\times 10^{7}$ yrs if no gas
replenishment from the arms occur.  If a ZAMS Salpeter IMF with an
upper (lower) mass cutoff of 100 M$_{\odot}$ (0.1 M$_{\odot}$) is
adopted then a total stellar mass over the central ring of $M_{*} =
5.4 \times 10^{5}$ M$_{\odot}$ is generated in the current burst.
These values correspond to star formation efficiencies, SFE =
$M_{*}/(M_{*} + M_{H_{2}})$, of $\sim$10\% over the central ring,
peaking at the nuclear $x_{1}$ - $x_{2}$ orbit intersections.  The SFE
drops to $\lsim$ 4\% along the molecular arms, similar to Galactic
disk values.

\subsection{Gas Inflow, Stability and Triggered Star Formation in 
Maffei 2 \label{gasinflow}}

What drives the star formation in the nucleus?  Is it the large
molecular gas surface density or is there evidence for a trigger that
is unique to the nuclear region?  What is responsible for the large
concentration of nuclear gas in Maffei 2?  With these observations we
can address the link between star formation and molecular gas on GMC
sizescales in the nuclear region of Maffei~2.

In the context of our dynamical model, the presence of the large gas
mass is probably due to slow inflow along the nuclear bar
\citep[eg.][]{RHV79,A92,TH92,RVT97,SVRTT05}.  Is there sufficient gas
inflow to produce the observed star formation?  It is assumed that the
inflowing gas will form stars, and that the star formation process is
initiated at the location of the $x_{1}$-$x_{2}$ orbit intersection in
the nuclear ring. Then the radial gas mass flux at this galactocentric
radius determines the star formation rate.  The gas mass flux is
related to the average inflow velocity, $v_{inf}$, the average arm
mass surface density, $\overline{\Sigma}_{arm}$, and the arm width,
$w$.  From the $^{13}$CO data $\overline{\Sigma}_{arm}~ \simeq ~210$
M$_{\odot}$ pc$^{-2}$ and $w \simeq 5^{''}$ (80 pc).  Inflow
velocities are determined from the bar model. Typical values are -20
to -40 km s$^{-1}$ along the bar arms.  Averaged over the arm area
only, $v_{r}~\simeq~$ -21 km s$^{-1}$.  Adopting these values (an
upper limit), a mass inflow rate, $\dot{M}_{inf} \sim 0.7$ M$_{\odot}$
yr$^{-1}$ is obtained.  Since $\dot{M}_{inf}$ is a factor of $\sim$5
larger than the nuclear ring star formation rate estimated from the
millimeter continuum, the inflow rate is sufficient to fuel the
nuclear starburst even with modest efficiency.

Does the molecular gas form stars due to gravitational instabilities
or is it directly triggered?  The gravitational stability of a thin,
rotating disk can be estimated from the Toomre $Q$ parameter
\citep[][]{S60,T64}.  A gas disk is unstable to gravitational collapse
if $Q ~= ~\kappa \sigma / \pi G \Sigma_{gas} <$ 1, where $\kappa$ is
the epicyclic frequency, $\sigma$ is the gas velocity dispersion and
$\Sigma_{gas}$ is the gas surface density.  Figure \ref{Fbarrot}b
displays the azimuthally averaged values of $\kappa$ from the bar
model together with the observed $^{13}$CO velocity dispersion, mass
surface density, and corresponding $Q$ values.  $Q$ is 8--10 across
the central ring region containing the starburst and remains $>$1 over
the central 30$^{''}$ radius.  Clearly the data are not consistent
with star formation occurring in $Q<1$ gravitationally unstable gas.
This is not surprising for gas in the very center of galaxies given
the (1) strong noncircular motions present, (2) the failure of the
thin differentially rotating disk approximation and (3) potentially
strong turbulence and magnetic fields \citep[e.g.,][]{E99,C01,WB02}.

Another estimator for gravitational stability that may be more
suitable to nuclear gas can be obtained from \citet[][]{E94}.
\citet[][]{E94} estimates the critical density above which gas in a
ring associated with an ILR can collapse to form stars as
$\rho_{crit}= 0.6 ~\kappa^{2}/G$, or $n_{crit}~ =~ 2.08 \times 10^{-3}
~\kappa^{2}$(km s$^{-1}$ kpc$^{-1}$).  The epicyclic frequency at the
radius of the ring is $\kappa ~\simeq ~ 2000$ km s$^{-1}$ kpc$^{-1}$
which implies ring densities must be $n_{H2} > ~ 8 \times 10^{3}
~cm^{-3}$ to be unstable to collapse.  From the LVG analysis we find
that the average density of the CO-emitting gas along the central ring
is an order of magnitude lower than this value.

A lower limit to the stability of the molecular clouds can be set by
assuming the clouds remain gravitationally bound against tidal forces.
A cloud of mass, $m$, will remain bound if $m ~\gsim~ M(R) (r/R)^3$,
where $M(R)$ is the total mass enclosed within a galactocentric
radius, $R$, and $r$ is the size of the molecular cloud
\citep[eg.,][]{SBGB91}.  Clouds with densities of $n_{H_{2}}^{tidal}
~\gsim~ 3.6 M_{M_{\odot}}(R) R_{pc}^{-3}$ remain bound.  For $R$ = 80
pc and $M_{M_{\odot}}(80 ~pc) ~=~ 1.1 \times 10^{8}$ M$_{\odot}$,
values applicable to Maffei 2's nuclear ring, $n_{H_{2}}^{tidal}
~\simeq ~ 630$ cm$^{-3}$.  This value is close to the densities
inferred from our LVG analysis.  The average molecular gas densities
in the central ring are too low to be gravitationally unstable, and
are likely only marginally tidally bound.  Indeed, along much of
the central molecular ring not associated with the sites where the
arms terminate, little star formation is observed.

It appears, then, that gravitational instability is not the answer.
Instead we consider the possibility that the star formation is
triggered by events external to the clouds.  Star formation in the
nucleus of Maffei 2 is concentrated at the location of the $x_1$-$x_2$
orbit intersections indicated by our modeling.  At these $x_1$-$x_2$
orbit intersection regions star formation appears to be triggered by
the collision of gas flowing inward along the arms of the bar with the
existing, more diffuse gas of the central ring.

We propose that the evolution of the nuclear starburst has proceeded
as follows.  A recent interaction between a small companion and Maffei
2 has driven a large quantity of gas into the nucleus, building up a
compact central bulge seen in the NIR \citep[][]{HMGT93a,HTH96}.
Assuming the potential generated by this compact bulge is slightly
oval (a few percent is all that is necessary), it has forced the
nuclear molecular gas into the bar distribution currently observed.
Inflow along the nuclear $x_{1}$ orbits piles up gas at the nuclear
$x_{1}-x_{2}$ orbit intersections.  The interaction results in a
fraction of the molecular gas going to the formation of dense cloud
cores which collapse and trigger the star formation events at GMC D
and just downstream of GMC F.  Gas not incorporated into the dense
component at these locations is then tidally sheared into the moderate
density, nearly uniformly distributed ring, which in turn becomes the
target for future collisions with infalling gas.  As long as there is
gas flowing inward the burst of star formation at the arm--ring
intersection can be sustained.

This scenario provides a good framework for all of the molecular gas
and millimeter continuum observed toward Maffei 2, with one exception,
the star formation near GMC E.  This star-forming complex is on the
$x_{2}$ orbit but not at either of the $x_{1}$-$x_{2}$ intersection
regions. Nor is it a strong HCN source (though some HCN emission is
seen).  Why is star formation occurring here?  Two possibilities come
to mind: (1) The star formation is triggered by the molecular gas
associated with GMC E interacting with the molecular gas towards GMC D
after having traversed one half of the $x_{2}$ orbit.  (2) The star
formation here reflects a slightly earlier epoch event associated with
its passing through the southern $x_{1}$ - $x_{2}$ interaction region.
It is now being seen with a time lag equal to the traversed portion of
the ring divided by the orbital velocity.  From the nuclear ring
parameters the time lag would be $\sim$ 1 Myr.  That the spectral
index of the millimeter continuum is somewhat steeper towards GMC E,
possibly suggesting a slightly older starburst with more evolved and
less dense \ion{H}{2} regions, may favor the latter scenario.

\subsection{Comparisons of Maffei 2 with Other Nearby Nuclei}

The nuclear morphology of Maffei 2 is similar to that observed in the
bright nuclei of the barred galaxies, IC 342, NGC 6946 and M 83
\citep[eg.][]{I90,RV95, SMPWI04} and the central molecular zone of the
Galaxy \citep[eg.][]{BGSBU91,RCMWA06}.  All have nuclear bar
morphologies reminiscent of their large scale analogs
\citep[e.g.,][]{A92}.  In IC 342 and M 83, it remains somewhat
ambiguous whether they are nuclear bars or just the inner portions of
the large scale bar, due to a combination of having massive clusters
that potentially influence the dynamics
\citep[e.g.,][]{SBM03,CTBM04,SMPWI04,SBMC07} and low inclination,
which hampers kinematic studies.  Maffei 2's kinematics leave little
doubt that it is a true double bar.  In fact the CO velocity field in
the nucleus of Maffei 2 is perhaps the best current example of nuclear
non-circular, bar motions outside our own Galactic Center.  Therefore
Maffei 2 can be added to NGC 6946 as confirmed double barred galaxies,
but with a physical scale about three times larger.  It is interesting
that like NGC 6946 \citep[and NGC 2974;][]{I90,EGF03,SBEL06}, our
CO(1--0) observations imply the existence of straight shocks in
nuclear bars.  The inner ring and offset straight shocks do not appear
to be common features of hydrodynamical models of secondary bars
\citep[eg.,][]{SH02,MTSS02}.  Moreover, this conclusion seems to hold
true for galaxies with both strong large scale bars (Maffei 2) and
weak large scale bars (eg. NGC 6946).

These nuclear bars also influence physical conditions of the gas.
Despite the presence of luminous star-forming complexes in these
nuclei, emission from the lines of the CO isotopomers is dominated by
subthermally excited emission from low excitation (T$_{ex} ~\sim 5 -
15$ K) gas, and that this emission represents the properties of the
bulk of the molecular gas.  While the ISM in Maffei 2 appears slightly
warmer than IC 342 \citep[][]{MT01} its T$_{k}$ are very close to the
average properties of NGC 6946 and the outer gas lanes of the Galactic
Center \citep[][]{HWBM93,PJBH98,MT04,NTKO07}.  However, densities in
the nucleus of Maffei 2 are consistently about 0.5 dex lower than NGC
6946 and the Galactic Center.  We suggest this comes from Maffei 2
having a stronger nuclear bar than the other two nuclei, resulting in
more dramatic disruption and redistribution of its nuclear ISM. 

\section{Summary}

New aperture synthesis maps are presented for emission in the J=2--1
and 1--0 transitions of $^{13}$CO and C$^{18}$O, as well as the J=1--0
lines of HCN and CO in the central arcminute ($\sim$ 1 kpc) of Maffei
2.  The H$_{2}$ column density as traced by optically thin CO
isotopologues is similar in morphology to what is implied from
$^{12}$CO, except that the emission from the isotopologues is more
closely confined to the two extended molecular arm ridges and more
uniformly distributed across the central ring.  The dense gas traced
by HCN(1-0) is more confined to the center of the galaxy than the CO
emitting gas.

The central molecular bar contains five main peaks that resolve into
at least 17 distinct GMCs, with radii of $\sim$40--110 pc and
linewidths $\gsim$ 40 km s$^{-1}$.  In the two innermost molecular
cloud complexes, at galactocentric radii of $\sim 5$\arcsec\ (80 pc
from the dynamical center), the GMCs are distinctly nonspherical,
elongated along the nuclear bar, with linewidths as large as 100 km
s$^{-1}$.  These GMCs are probably being tidally stretched due to the
nuclear potential.

The H$_{2}$ column density for the central GMCs is $N_{H_{2}}~\simeq
~4.4 - 10\times 10^{22}~ cm^{-2}$ ($\overline{\Sigma} ~\sim 950 -
2200$ M$_{\odot}$ pc$^{-2}$), corresponding to mean optical
extinctions of $A_v \sim$40--100.  The molecular mass within the
central 20\arcsec\ galactocentric radius ($\sim$300 pc) is
2.1$\times10^{7}~M_{\odot}$, while the dynamical mass in the same
region is M$_{dyn}(20^{''})=7.3\times 10^{8}~M_{\odot}$.  The
molecular mass is only a few percent of the dynamical mass.
Excitation temperatures, assuming $^{13}\tau (^{18}\tau) ~\sim$ 1
~($\ll$1), are T$_{ex}~\sim ~ 3~-~6$ K over much of the central 500 pc
for both $^{13}$CO and C$^{18}$O.  These T$_{ex}$ values are low
compared with the brightness temperature observed in CO ($\gsim 30$ K)
indicating subthermal excitation, and that the average densities of
the GMCs are probably only moderate.  Single component LVG analysis of
the GMCs in CO, $^{13}$CO, and C$^{18}$O yield best-fit solutions of
$n_{H_{2}}~\simeq ~ 10^{2.75}~ cm^{-3}$ and T$_{kin} ~\simeq ~ 20 -
30$ K.  Average densities estimated from the total C$^{18}$O column
densities are consistent with these values.

The $^{13}$CO and C$^{18}$O lines are weaker than expected from
CO(1-0), which appears to be overluminous per unit gas mass across the
starburst region.  Column densities derived from both C$^{18}$O and
1.4 mm dust continuum emission imply that $X_{CO}$(Maf~2) is about
2--4 times lower than the Galactic value, similar to $X_{CO}$ values
found for the centers of other large spirals, including our own.  The
weakness of the isotopologues at large galactocentric radii and in the
``off-arm'' spray regions of Maffei 2, suggest that in these regions
either the isotopologues cease to effectively trace molecular gas or
that the Galactic conversion factor overestimates the molecular
column.  The lack of applicability of the Galactic $X_{CO}$ to the
clouds in the center of Maffei 2 is probably due to the effect of bar
motions and strong tides on the structure and dynamics of these
clouds.

Millimeter continuum emission reveals three prominent locations of
star formation with the most intense occurring where the molecular bar
intersects the nuclear ring.  Lyman continuum rates of N$_{Lyc}\sim
3$--$5 \times 10^{51}$ s$^{-1}$ are implied for individual regions.
The total rate for the entire nucleus is $_{Lyc}\sim 2.6 \times
10^{52}$ s$^{-1}$, or SFR $\sim$0.26 M$_{\odot}$ yr$^{-1}$.  

A P-V diagram of the nucleus of Maffei 2 shows a distinct
``parallelogram'' indicating molecular gas response to a barred
potential.  The morphological and kinematic data confirms Maffei 2 as
true double barred galaxy.  We suggest a bar model where the nuclear
gas distribution and velocity is governed by a small nuclear bar of $r
= \sim$110 pc.  An upper limit to the mass inflow rate along the
nuclear bar is $dM/dt~ \lsim ~0.7$ M$_{\odot}$ yr$^{-1}$, enough to
drive the current star formation rate seen at the end of the bar arms
and populate the nuclear ring with gas.  The locations of star
formation and the dense gas in the central region appear to coincide
with the location of the $x_{1}-x_{2}$ orbit crossings of the nuclear
bar, consistent with dynamical triggering of the the star formation.

\acknowledgements 

We are grateful to the faculty and staffs of OVRO and BIMA for their
support during the observations.  We thank the referee, Marshall
McCall, for a careful and insightful reading of the paper.  DSM
acknowledges support from NSF AST-0506669, the Laboratory for
Astronomical Imaging at the University of Illinois (NSF AST-0228953)
and NRAO.  The National Radio Astronomy Observatory is a facility of
the National Science Foundation operated under cooperative agreement
by Associated Universities, Inc.  Additional support for this work is
provided by NSF grant AST-0071276 and AST-0506469 to JLT.

\clearpage

\begin{deluxetable}{lcc}
\tablenum{1}
\tablewidth{0pt}
\tablecaption{Maffei 2 Basic Data}
\tablehead{
\colhead{Characteristic}
&\colhead{Value}
&\colhead{Reference}
}
\startdata 
Revised Hubble Class      &SBb(s) pec   & 1  \\
Dynamical Center        & $\alpha(J2000) = 02^{h} 41^{m} 54^{s}.90\pm 
0.^{s}15~~~$    &3   \\
  &$ \delta(J2000) = +59^{o} 36' 14.^{''}4\pm 2^{''}$   &   \\
$\ell^{II}$,b$^{II}$ & 136.5$^{o}$,-0.3$^{o}$  &1   \\
V$_{lsr}$   &-30 kms$^{-1}$   &3   \\
Adopted Distance  & 3.3 Mpc  &4   \\
Inclination Angle  & 67$^{o}$  &3   \\
Position Angle  &206$^{o}$   &3   \\
M(HI)\tablenotemark{a}&$4.0 \times 10^{8}~M_{\odot}$ &3 \\
M(H$_{2}$)\tablenotemark{b}&$8.5 \times 10^{8}~M_{\odot}$ &5 \\
\enddata
\tablenotetext{a}{Corrected for adopted distance.}
\tablenotetext{a}{Corrected for adopted distance and 
assumed CO conversion factor.}
\tablerefs{(1) \citet[][]{HMGT93a}; (2) \citet[][]{HT91}; 
(3) \citet[][]{HTH96}; (4) \citet[][]{FLMR07}, See text; 
(5) \citet[][]{MW04}.} 
\label{Tmaf}
\end{deluxetable}

\clearpage

\begin{deluxetable}{lccccccc}
\tablenum{2}
\tablewidth{0pt}
\tablecaption{Observational Data}
\tablehead{
\colhead{Transition}
&\colhead{$\nu_{o}$}
&\colhead{T$_{sys}$}
&\colhead{$\Delta V_{chan}$}
&\colhead{$\nu_{band}$}
&\colhead{Beamsize}
&\colhead{Noise}
&\colhead{Det.\tablenotemark{a}}
\\
\colhead{}
&\colhead{}
&\colhead{}
&\colhead{}
&\colhead{}
&\colhead{}
&\colhead{Level}
&\colhead{Flux}
\\
\colhead{}
&\colhead{(GHz)}
&\colhead{(K)}
&\colhead{\footnotesize (km s$^{-1}$)}
&\colhead{\footnotesize (MHz)}
&\colhead{\footnotesize ('';$^{o}$)}
&\colhead{\footnotesize (mK/mJy/Bm)}
&\colhead{\footnotesize (\%)}
}
\startdata
OVRO: &&&&&& \\
  HCN(1-0)\tablenotemark{b}& 88.63 & 300-410 & 13.53& 128 
&3.8x3.3;-29$^{o}$ & 120/10 & 65 \\
  $^{13}$CO(1-0)\tablenotemark{b}& 110.20 &230-430 &2.72
&128 &3.9x3.4;-76$^{o}$& 77/10 & 92\\
  $^{13}$CO(2-1)\tablenotemark{d}& 220.40 & 500-1000 & 2.72
& 128 &3.3x2.9;-76$^{o}$ & 75/28 & $\sim$50\\
  C$^{18}$O(1-0)\tablenotemark{b}& 109.78 & 240-430 & 10.92& 128 
&2.6x2.2;-84$^{o}$ & 130/7.5 &\nodata\\
  C$^{18}$O(2-1)\tablenotemark{b} & 219.56 &300-1000 &5.46 
&128 &1.7x1.5;-62$^{o}$& 150/32 &\nodata \\
  3.4 mm\tablenotemark{b} & 88.92   &300-410 &\nodata  &1000 
&2.5x2.5;0$^{o}$ & 17/0.67 &\nodata \\
  2.7 mm\tablenotemark{c} & 109.5   &230-430 &\nodata  &1000 
&3.9x3.4;-76$^{o}$ & 3.9/0.50 & \nodata \\
  1.4 mm\tablenotemark{b} & 219.3   &300-1000 &\nodata  &1000 
&1.7x1.5;-62$^{o}$ & 25/2.5 & \nodata \\
BIMA: &&&&&& \\
  $^{12}$CO(1-0)\tablenotemark{e}& 115.27 &380-1300 &4.07 &172 & 
3.2x3.1;-14$^{o}$& 1400/0.15 & 105\\
\enddata
\tablecomments{Dates for the observations are 
$^{13}$CO(1--0), 1994 October 23--1995 January 2;   
$^{13}$CO(2--1), 1993 October 26--1994 January 13;
C$^{18}$O(1--0) and C$^{18}$O(2--1), 1998
October 19--1999 January 5;
HCN(1--0), 1999 January 28--1999 March 29;
$^{12}$CO(1-0), 2004 March 15.} 
\tablenotetext{a}{The percentage of the single-dish flux detected 
by the interferometers.  Single-dish integrated intensities 
come from the following: CO(1-0) and  $^{13}$CO(1-0), 
\citep[][]{WCC88}, HCN(1-0) \citep[][]{RJHTM92} and $^{13}$CO(2-1), 
\citep[][]{WHEGGJRS92}.}
\tablenotetext{b}{Phase Center \#1: V$_{LSR}$=-30 km ~s$^{-1}$
$\alpha=02^{h}38^{m}08^{s}.00$, $\delta=+59^{o}23'20.^{''}0$ (B1950),
\#2: $\alpha=02^{h}38^{m}08^{s}.25$, $\delta=+59^{o}23'27.^{''}0$
(B1950). } 
\tablenotetext{c}{Phase Center \#1: V$_{LSR}$=-28 $km ~s^{-1}$
$\alpha=02^{h}38^{m}07^{s}.00$, $\delta=+59^{o} 23' 33.^{''}0$ (B1950)
\#2: $\alpha=02^{h}38^{m}09^{s}.00$, $\delta=+59^{o}23'40.^{''}0$
(B1950).} 
\tablenotetext{d}{Phase Center \#1: V$_{LSR}$=-28 
$km ~s^{-1}$ $\alpha=02^{h}38^{m}07^{s}.50$, $\delta=+59^{o}23'08.^{''}0$ 
(B1950) \#2: $\alpha=02^{h}38^{m}08^{s}.80$, $\delta=+59^{o}23'33.^{''}0$ 
(B1950)} 
\tablenotetext{e}{Phase Center \#1: V$_{LSR}$=-15 $km ~s^{-1}$
$\alpha=02^{h}41^{m}59^{s}.19$, $\delta=+59^{o}36'46.^{''}8$ (J2000),
\#2: $\alpha=02^{h}41^{m}55^{s}.0$, $\delta=+59^{o}36'15.^{''}0$
(J2000), \#3: $\alpha=02^{h}41^{m}50^{s}.31$, $\delta=+59^{o}35'43.^{''}2$
(J2000).} 
\label{Tobs}
\end{deluxetable}

\clearpage

\begin{deluxetable}{lcccccc}
\tablenum{3}
\tablewidth{0pt}
\tablecaption{Measured Intensities}
\tablehead{
\colhead{}
&\colhead{$^{12}$CO(1-0)}
&\colhead{HCN(1-0)}
&\colhead{$^{13}$CO(1-0)}
&\colhead{$^{13}$CO(2-1)}
&\colhead{C$^{18}$O(1-0)}
&\colhead{C$^{18}$O(2-1)}
\\
\colhead{}
&\colhead{\footnotesize ($K~km~s^{-1}$)}
&\colhead{\footnotesize ($K~km~s^{-1}$)}
&\colhead{\footnotesize ($K~km~s^{-1}$)}
&\colhead{\footnotesize ($K~km~s^{-1}$)}
&\colhead{\footnotesize ($K~km~s^{-1}$)}
&\colhead{\footnotesize ($K~km~s^{-1}$)}
}
\startdata
A & $520\pm50$ & $\lsim 8.3$ & $8.5\pm0.9$ & \nodata & $< 5.4$ 
& \nodata \\
B & $540\pm50$ & $26\pm3$ & $38\pm4$ & $\sim 4.8$ & $\sim 8.7$ 
& $\sim 5.2$ \\
C & $780\pm80$ & $56\pm6$ & $56\pm6$ &$9.3\pm1.8$ & $23\pm2$ 
& $13\pm3$ \\
D & $1300\pm100$ & $110\pm10$ & $93\pm9$ & $73\pm10$ & $26\pm3$ 
& $18\pm4$ \\
E & $1200\pm100$ & $100\pm10$ & $120\pm10$ & $83\pm20$ & $28\pm3$ 
& $21\pm4$ \\
F & $680\pm70$ & $83\pm8$ & $85\pm9$ & $41\pm8$ & $21\pm2$ 
& $17\pm3$ \\
G & $530\pm50$ & $23\pm2$ & $63\pm6$ & $11\pm2$ & $17\pm2$ 
& $\lsim 3.5$  \\
H & $170\pm20$ & $<8.3$ & $\lsim 6.8$ & \nodata   & $< 5.4$ 
&\nodata  \\
\enddata
\tablecomments{Uncertainties are the larger of the 
map uncertainty or the absolute calibration uncertainties 
(assuming to be 10 \% for the 3 mm lines and 20 \% 
for the 1 mm lines).  Refer to Figure \ref{Fgmc} or Table 
\ref{Tgmc} for the GMC positions.  Values are obtained from 
maps convolved to the $^{13}$CO(1-0) beamsize.  For 
these lower resolution maps the GMCs are sampled at the ``1'' 
component of each GMC (except G which is sampled at G3), which 
corresponds approximately to the peak in the lower resolution 
maps.}
\label{Tint}
\end{deluxetable}

\clearpage

\begin{deluxetable}{lcccccc}
\tablenum{4}
\tablewidth{0pt}
\tablecaption{Giant Molecular Clouds}
\tablehead{
\colhead{GMC}
&\colhead{$\alpha, ~\delta$}
&\colhead{$a \times b;~ pa$}
&\colhead{$\Delta v_{1/2}$}
&\colhead{$v_{o}$}
&\colhead{T$_{pk}$}
&\colhead{$M_{vir}$}
\\
\colhead{}
&\colhead{(02$^{h}$41$^{m}$;59$^{o}$36$^{'}$) }
&\colhead{($pc \times pc;^{o}$)}
&\colhead{($km~s^{-1}$)}
&\colhead{($km~s^{-1}$)}
&\colhead{$(K)$}
&\colhead{($10^{6}~M_{\odot}$)}
}
\startdata
A1 &57.17,42.4 &$51\times 36$;46  & 78$\pm$3& -66$\pm$1& 13&34  \\
A2 &57.86,42.2 &$29\times 29$;120  & 82$\pm$8 &-59$\pm$3  & 9&30  \\
A3 &57.51,41.6 &$52\times 31$;23  & 95$\pm$7 &-67$\pm$3  & 21&47  \\
B1  &55.95,26.9 &$76\times 42$;180  & 75$\pm$2& -89$\pm$1 &9 & 42   \\
B2  &55.75,29.6 &$55\times 41$;18  & 41$\pm$2& -80$\pm$1 &17 & 10   \\
C  &55.48,24.8 &$110\times 30$;26  & 61$\pm$1& -84$\pm$1 & 18& 29   \\
D1 &55.14,20.7 &$76\times 33$;28  & 84$\pm$1& -74$\pm$1 &31 & 46  \\
D2 &54.78,24.7 &$39\times 23$;34  & 8.6$\pm$3& -54$\pm$1 &9  & 0.29  \\
E &55.08,18.4 &$110\times 26$;160\tablenotemark{b}  & 110$\pm$1 & -79$\pm$1 
& 19 & 87  \\
F  &54.86,09.6 &$95\times 33$;36  & 51$\pm$1& -2.0$\pm$1 &24 & 19  \\
G1  &54.39,03.2 &$65\times 29$;36  & 47$\pm$1& 20$\pm$1 & 17 & 12  \\
G2  &54.34,00.1 &$44\times <17$;55 & 69$\pm$4 & 38$\pm$2 & 9 &$<$18  \\
G3  &54.12,59.6\tablenotemark{a} &$69\times 29$;73  & 66$\pm$2& 38$\pm$2 
& 9 & 25  \\
G4  &53.88,57.9\tablenotemark{a} &$<19\times <17$;--  & 79$\pm$7& 30$\pm$3 
& 12 &$<$14  \\
G5  &53.77,00.7 &$73\times 22$;35  & 64$\pm$2& 23$\pm$1 & 14 &22  \\
H1  &53.76,04.7 &$64\times 23$;57& 18$\pm$1& 14$\pm$1 & 12& 1.7 \\
H2  &53.52,01.4 &$63\times \lsim 17$;150& 70$\pm$3 & 24$\pm$2 & 9 
&$\lsim$21 \\
\enddata
\tablecomments{Positions are based on fitting the uniformly 
weighted CO(1-0) data.  Refer to Figure \ref{Fgmc} for the 
locations of each GMC.  Uncertainties are 1$\sigma$ from the 
least-squares Gaussian fits to the data.  A GMC is considered unresolved 
if its deconvolved size is less than 1/2 of the beam minor axis.}  
\tablenotetext{a}{02$^{h}$41$^{m}$; 59$^{o}$35$^{'}$.}  
\tablenotetext{b}{Size is uncertain due to blending.}  
\label{Tgmc}
\end{deluxetable}

\clearpage

\begin{deluxetable}{lcccccc}
\tablenum{5}
\tablewidth{0pt}
\tablecaption{Millimeter/Radio Continuum Flux and Spectral Indices}
\tablehead{
\colhead{Parameter}
&\colhead{I}
&\colhead{II}
&\colhead{III}
&\colhead{IV}
&\colhead{V}
&\colhead{Total}
}
\startdata
($\alpha_{o},\delta_{o}$) & 55.09;21.0 & 55.10;17.6 & 55.14;16.0 & 
55.17;13.5 & 54.54;01.8 & \nodata \\
\hline 
$S_{mJy}(4.885)$ &14.3$\pm$0.7 &19.4$\pm$1 &15.2$\pm$0.8 &11.4$\pm$0.6 &
2.7$\pm$0.1 & 97$\pm$5 \\
$S_{mJy}(14.96)$ &7.0$\pm$0.4 & 9.6$\pm$0.5 &7.0$\pm$0.4 & 5.6$\pm$0.3 &
1.0$\pm$0.2 & 48$\pm$2 \\
$S_{mJy}(89.96)$\tablenotemark{a} &5.8$\pm$0.9 & 5.6$\pm$0.9 &
$\lsim$3.9 & 5.0$\pm$0.8 & 1.5$\pm$0.6 &32$\pm$5 \\
$S_{mJy}(110.2)$\tablenotemark{b} &7.0$\pm$1 &$\sim$7.0\tablenotemark{c} & 
4.7$\pm$0.7 & 4.1$\pm$0.7 &2.7$\pm$0.5 & 26$\pm$3  \\
$S_{mJy}(219.3)$ &19$\pm$4 &21$\pm$4 &10$\pm$3 &$\sim$5.3 &9.7$\pm$3 & 
79$\pm$16\\
\hline
$\alpha^{6}_{2}$ & -0.6$\pm$0.1 & -0.6$\pm$0.1 & -0.7$\pm$0.1 &   
-0.6$\pm$0.1 & -2.0$\pm$0.2 & -0.6$\pm$0.1  \\
$\alpha^{2}_{3.4}$ & -0.1$\pm$0.1 & -0.3$\pm$0.1 & $<-0.33$ & 
-0.1$\pm$0.1 & 0.2$\pm$0.3 & -0.2$\pm$0.1  \\
$\alpha^{2.7}_{1.4}$ & 1.5$\pm$0.5 & $\sim 1.6$ & 1.1$\pm$0.6 & 
$\sim 0.4$ & 1.9$\pm$0.6 & 2.0$\pm$0.5  \\
\enddata
\tablecomments{Uncertainties are based on the larger 
of the map noise or 5\% for 6 and 2 cm (Turner \& Ho 1994), 15 \% 
absolute calibration at 3 mm (larger uncertainties reflect possible errors 
in line removal), and 20 \% at 1 mm. Upper limits are 2$\sigma$.} 
\tablenotetext{a}{Continuum emission is corrected 
for HCN(1-0) \& HCO$^{+}$(1-0) line flux.} 
\tablenotetext{b}{Continuum emission is corrected 
for $^{13}$CO(1-0) line flux.} 
\tablenotetext{c}{Uncertain due to confusion with I.} 
\label{Tcont}
\end{deluxetable}

\clearpage

\begin{deluxetable}{lccccc}
\tablenum{6}
\tablewidth{0pt}
\tablecaption{Star Formation Rates}
\tablehead{
\colhead{Source}
&\colhead{$^{ff}S_{mJy}$}
&\colhead{$N_{Lyc}$}
&\colhead{SFR}
&\colhead{$^{D}S_{mJy}$}
&\colhead{$^{D}M_{H_{2}}$} \\
\colhead{(1)}
&\colhead{(2)}
&\colhead{(3)}
&\colhead{(4)}
&\colhead{(5)}
&\colhead{(6)}
}
\startdata
I &4$\pm$0.6 & 5$\pm$0.6 &0.05$\pm$0.01 &15$\pm$3 & 4.8$\pm$0.8 \\
II &4$\pm$0.6 &5$\pm$0.6 &0.05$\pm$0.01 &19$\pm$4 & 6.1$\pm$1.0 \\
III &2$\pm$0.6 &3$\pm$0.6 &0.03$\pm$0.01 &7$\pm$3 & 2.3$\pm$0.8 \\
IV &3$\pm$0.6 &4$\pm$0.6 &0.04$\pm$0.01 &$<2.5$  &$<0.81$ \\
V &$<$1.2  & $<$1.6 & $<$0.02&  10$\pm$3 & 3.3$\pm$0.8 \\
Total & 20$\pm$3& 26$\pm$3&0.26$\pm$0.05 & 50$\pm$10& 16$\pm$2.6\\
\enddata
\tablecomments{Table columns: (1) The radio continuum sources listed 
in Table \ref{Tcont}. (2) Free-free flux, $^{ff}S_{mJy}$,
at 88.92 GHz, is derived from the spectral fits (Figure \ref{Fseds}).
(3) N$_{lyc}$ is the number of ionizing photons in units of
$10^{51}~s^{-1}$, or $\sim 100$ O7 stars \citep[][]{VGS96} implied by
$^{ff}S_{mJy}$. (4) SFR is the star formation rate in units of
$M_{\odot}~yr^{-1}$, based on the conversion from N$_{Lyc}$ to SFR of
Kennicutt (1998). (5) Thermal dust flux, $^{D}S_{mJy}$ at 219.3 GHz is
derived from the spectral fits (Figure \ref{Fseds}). (6)
$^{D}M_{H_{2}}$ is the total mass in units of $10^{6}~M_{\odot}$
implied from the dust flux, $^{D}S_{mJy}$, assuming a gas-to-dust
ratio of 100.}
\label{Tstarf}
\end{deluxetable}

\clearpage

\begin{deluxetable}{lcccccc}
\tablenum{7}
\tablewidth{0pt}
\tablecaption{Observed Line Ratios}
\tablehead{
\colhead{GMC}
&\colhead{$\rm \frac{^{12}CO(1-0)}{^{13}CO(1-0)}$}
&\colhead{$\rm \frac{^{12}CO(1-0)}{C^{18}O(1-0)}$}
&\colhead{$\rm \frac{^{13}CO(1-0)}{C^{18}O(1-0)}$}
&\colhead{$\rm \frac{^{12}CO(1-0)}{HCN(1-0)}$}
&\colhead{$\rm \frac{^{13}CO(2-1)}{^{13}CO(1-0)}$}
&\colhead{$\rm \frac{C^{18}O(2-1)}{C^{18}O(1-0)}$}
}
\startdata
A &$62\pm$20 & $> 96$ & $>1.6$ & $>63$ &\nodata  
& \nodata \\
B &$14\pm$2 & $\sim 95$ & $\sim 6.7$ & $21\pm$4 & $\sim$0.24 
& $\sim$0.90 \\
C &$14\pm$2 & $34\pm$7 & $2.5\pm$0.4 &$14\pm$2 & $0.18\pm$0.05 
& $0.56\pm$0.1 \\
D &$14\pm$2 & $50\pm$6 & $3.7\pm$0.5 & $12\pm$2  & $0.77\pm$0.2 
& $0.68\pm$0.2 \\
E &$10\pm$1 & $41\pm$4 & $4.0\pm$0.5 & $12\pm$2 & $0.75\pm$0.2 
& $0.78\pm$0.2 \\
F &$7.8\pm$1 &$32\pm$5 & $4.1\pm$0.7 & $8.1\pm$1 & $0.51\pm$0.1 
& $0.79\pm$0.2 \\
G &$8.4\pm$1 & $31\pm$6 & $3.8\pm$0.9 & $23\pm$5 & $0.28\pm$0.07 
& $\lsim0.82$  \\
H &$\sim 25$ &$> 31$ & $>1.3$ & $>20$   & \nodata & \nodata  \\
\enddata
\tablecomments{Based on the resolution of the $^{13}$CO(1-0) data.  
No corrections for differences in resolved out flux have been 
included.}
\label{Tratio}
\end{deluxetable}

\clearpage

\begin{deluxetable}{lcccccccccc}
\rotate
\tablenum{8}
\tablewidth{0pt}
\tablecaption{CO Conversion Factor}
\tablehead{
\colhead{}
&\colhead{$^{13}$T$_{ex}$}
&\colhead{$^{13}N_{H_{2}}$}
&\colhead{$^{18}$T$_{ex}$}
&\colhead{$^{18}N_{H_{2}}$}
&\colhead{$^{Xco}N_{H_{2}}$}
&\colhead{$^{D}N_{H_{2}}$}
&\colhead{$^{V}N_{H_{2}}$}
&\colhead{$^{L}N_{H_{2}}$}
&\colhead{$\frac{^{MW}X_{CO}}{^{13}X_{co}}$}
&\colhead{$\frac{^{MW}X_{CO}}{^{18}X_{co}}$} 
\\
\colhead{(1)}
&\colhead{(2)}
&\colhead{(3)}
&\colhead{(4)}
&\colhead{(5)}
&\colhead{(6)}
&\colhead{(7)}
&\colhead{(8)}
&\colhead{(9)}
&\colhead{(10)}
&\colhead{(11)} 
}
\startdata
A & [5.0] &$\sim 0.77$ & [5.0] & $<1.3$ &10$\pm$1.0 & \nodata &  130 &  \nodata 
& $\sim$13 & $>7.7$ \\
B &$3.1\pm1.0$ & $4.4\pm0.5$ & $\sim 6.7$ &$\sim$2.0 &11$\pm$1.0 &\nodata  
& 51 & \nodata & $2.5\pm0.3$ & $\sim$5.5 \\
C &$2.9\pm0.9$ & $7.7\pm 0.8$ & $4.8\pm1.6$ & $5.5\pm 0.6$ & 16$\pm$1.6 
&\nodata  &45 & 11 & $2.1\pm0.3$ & $2.9\pm 0.4$  \\
D & $6.1\pm2.3$ & $8.2\pm0.8$ &  $5.4\pm1.8$ & $6.1\pm0.6$  &26$\pm$2.6 
& $8\pm 2$ & 71 & 11 & $3.2\pm 0.4$ & $4.3\pm0.6$ \\
E &$6.0\pm2.3$ & $10\pm1.0$ & $5.9\pm2.0$ & $6.5\pm0.7$ & 24$\pm$2.4
& 11$\pm$3 & 150 & 14 & $2.4\pm0.3$ & $3.5\pm0.5$ \\
F & $4.5\pm1.5$ & $7.9\pm0.8$ & $6.0\pm2.0$ & $4.9\pm0.5$ & 14$\pm$1.4 
& $<$1.4 &47 & 14 & $1.8\pm0.2$ & $2.9\pm0.4$ \\
G &$3.4\pm1.1$ & $6.8\pm0.7$ & $\lsim 6.2$ & $\lsim 4.0$ &10$\pm$1.0 
&5.6$\pm$2 & $>75$ & 11 & $1.5\pm0.2$ &$\gsim 2.5$ \\
H & [5.0] &$\lsim 0.61$ & [5.0] & $< 1.3$ & 3.4$\pm$0.3 &\nodata & $\lsim53$ 
& \nodata & $\gsim5.6$ & $>2.6$ \\
\enddata
\tablecomments{Column densities are in units of $10^{22}$ cm$^{-2}$
and have been corrected for resolved flux assuming the emission is
uniformly distributed over 30$^{''}$ and temperatures are in units of
K.  Uncertainties reflect only statistical uncertainties in the
intensities and do not include (larger) systematic uncertainties.
Table columns: (1) the molecular peaks; (2) the excitation temperature
derived from the $^{13}$C)(2--1)/$^{13}$CO(1--0) line ratio, assuming
$\tau_{13CO(1-0)} \simeq 1$ (\S \ref{exotau}); (3) the H$_{2}$ column
density, N$_{H_{2}}$, derived from the $^{13}$CO(1--0) line intensity
assuming $^{13}$T$_{ex}$ and $\tau_{13CO(1-0)} \simeq 1$; (4) and (5)
as in (2) and (3) except from C$^{18}$O with $\tau_{C18O(1-0)} \ll 1$;
(6) N$_{H_{2}}$ derived assuming a Galactic conversion factor of $2
\times 10^{20} (K km s^{-1})^{-1} cm^{-2}$
\citep[eg.,][]{Set88,Het97,DHT01}; (7) N$_{H_{2}}$ estimated by
averaging the gas mass estimated from dust emission in Table
\ref{Tstarf} over the 1.4 mm beamsize (Figure \ref{Fcont}); (8)
N$_{H_{2}}$ estimated by averaging the virial mass over the cloud
area, $\Sigma_{i} ^{V}M_{i}/\Sigma_{i} A_{i}$ for each cloud component
(Table \ref{Tgmc}); (9) N$_{H_{2}}$ estimated by averaging the LVG
derived density over the beamsize, assuming a third dimension of
$\sqrt{\theta_{a}\theta_{b}}$; (10) the ratio of N$_{H_{2}}$
calculated from the Galactic conversion factor to that from $^{13}$CO;
(11) the ratio of N$_{H_{2}}$ calculated from the Galactic conversion
factor to that from C$^{18}$O.  }

\label{Tx_co}
\end{deluxetable}

\clearpage

\begin{deluxetable}{lc}
\tablenum{9}
\tablewidth{0pt}
\tablecaption{Maffei 2 Bar Model Parameters}
\tablehead{
\colhead{Characteristic}
&\colhead{Value}
}
\startdata 
 position angle ....................... & 29$^{o}$\\
 $i$ & 67$^{o}$\\
 $\theta_{main}$\tablenotemark{a} & -17$^{o}$\\
 $\Omega_{main}$ & 40 km  s$^{-1}$ kpc$^{-1}$\\
 $\Omega_{nuc}$ & 135 km  s$^{-1}$ kpc$^{-1}$\\
 $\theta_{nuc}$\tablenotemark{b} & 7$^{o}$\\
 $^{main}r_{max}$\tablenotemark{c} & 3.6 kpc \\
 $^{main}v_{max}$\tablenotemark{c} & 172 km s$^{-1}$\\ 
 $^{nuc}r_{max}$\tablenotemark{c} & 155 pc \\
 $^{nuc}v_{max}$\tablenotemark{c} & 90 km s$^{-1}$\\ 
 n$_{main}$\tablenotemark{c} & 1.0 \\
 n$_{nuc}$\tablenotemark{c} & 3.25 \\
 $\epsilon_{main}$\tablenotemark{d} & 0.125 \\
 $\epsilon_{nuc}$\tablenotemark{d} & 0.075\\
 $\lambda$\tablenotemark{e} & 0.125\\
 $\mu$\tablenotemark{e} & 0.1875\\
\enddata
\tablenotetext{a}{Angle between the main bar and the major axis.}
\tablenotetext{b}{Angle between the main bar and the nuclear bar.}
\tablenotetext{c}{Brant model parameters for the two axisymmetric 
potentials.  The bar core radii are $\sqrt{2}$ smaller.}
\tablenotetext{d}{Strength of each bar \citep[see][]{W94}.}
\tablenotetext{e}{Magnitude of the radial ($\lambda$) and 
azimuthal ($\mu$) damping term  \citep[see][]{W94,MS97}.}
\label{Tbar}
\end{deluxetable}

\clearpage

\clearpage

\clearpage
\thispagestyle{empty}
\setlength{\voffset}{-18mm}

\clearpage
\setlength{\voffset}{0mm}

\clearpage

\clearpage

\clearpage

\clearpage

\clearpage

\clearpage

\clearpage

\clearpage

\clearpage

\end{document}